\documentclass{emulateapj}

\newcommand{\chandra}{{\it Chandra}}
\newcommand{\swift}{{\it Swift}}
\newcommand{\xmm}{{\it XMM-Newton}}

\begin{document}

\title{High-resolution X-ray spectroscopy of the evolving shock in the
2006 outburst of RS\,Ophiuchi}

\author{J.-U. Ness\altaffilmark{1}, J.J. Drake\altaffilmark{2},
S. Starrfield\altaffilmark{1},
M.F. Bode\altaffilmark{3}, T.J. O'Brien\altaffilmark{4},
A. Evans\altaffilmark{5}, S.P.S. Eyres\altaffilmark{6},
L.A. Helton\altaffilmark{7},
J.P. Osborne\altaffilmark{8}, K.L. Page\altaffilmark{8},
C. Schneider\altaffilmark{9}, C.E. Woodward\altaffilmark{7}
}

\altaffiltext{1}{School of Earth and Space Exploration, Arizona
State University, Tempe, AZ 85287-1404, USA: Jan-Uwe.Ness@asu.edu}
\altaffiltext{2}{Harvard-Smithsonian Center for Astrophysics, 60
Garden Street, Cambridge, MA 02138, USA}
\altaffiltext{3}{Astrophysics Research Institute, Liverpool John Moores University, Birkenhead, CH41 1LD, UK}
\altaffiltext{4}{Jodrell Bank Observatory, School of Physics \& Astronomy, University of Manchester, Macclesfield, SK11 9DL, UK}
\altaffiltext{5}{Astrophysics Group, Keele University, Keele, Staffordshire, ST5 5BG, UK}
\altaffiltext{6}{Centre for Astrophysics,
School of Computing, Engineering \& Physical Sciences,
University of Central Lancashire,
Preston, PR1 2HE, UK}
\altaffiltext{7}{Department of Astronomy, School of Physics \& Astronomy, 116 Church Street S.E., University of Minnesota, Minneapolis, MN 55455, USA}
\altaffiltext{8}{Department of Physics \& Astronomy, University of Leicester, Leicester, LE1 7RH, UK}
\altaffiltext{9}{Hamburger Sternwarte, Gojenbergsweg 112, 21029
Hamburg, Germany}

\begin{abstract}
 The evolution of the 2006 outburst of the recurrent nova
RS Ophiuchi was followed with 12 X-ray grating observations with
\chandra\ and \xmm. We present detailed spectral analyses
using two independent approaches. From the best dataset,
taken on day 13.8 after outburst, we reconstruct the
temperature distribution and derive elemental abundances.
We find evidence for at least two distinct temperature
components on day 13.8 and a reduction of temperature with
time. The X-ray flux decreases as a power-law, and the
power-law index changes from $-5/3$ to $-8/3$ around day
70 after outburst. This can be explained by different decay
mechanisms for the hot and cool components. The decay of the
hot component and the decrease in temperature are consistent
with radiative cooling, while the decay of the cool component
can be explained by the expansion of the ejecta.
We find overabundances of N and of $\alpha$ elements, which
could either represent the composition of the secondary that
provides the accreted material or that of the ejecta.
The N overabundance indicates CNO-cycled material. From
comparisons to abundances for the secondary taken from the
literature, we conclude that 20-40\% of the observed nitrogen
could originate from the outburst. The overabundance of the
$\alpha$ elements is not typical for stars of the spectral
type of the secondary in the RS\,Oph system, and white dwarf
material might have been mixed into the ejecta. However, no
direct measurements of the $\alpha$ elements in the secondary
are available, and the continuous accretion may have changed
the observable surface composition.
\end{abstract}


\section{Introduction}

 Nova explosions occur in binary systems containing a white dwarf
(WD) that accretes hydrogen-rich material from its companion.
When $10^{-6}-10^{-4}$\,M$_\odot$ have been accreted (depending
on the WD mass), ignition conditions for explosive nuclear
burning are reached and a thermonuclear runaway (TNR) occurs
\citep{st08}. Material dredged up from below the WD surface is
mixed with the accreted material and violently ejected. While
nuclear burning continues, the WD is surrounded by a pseudo
atmosphere, and the peak of the spectral energy distribution
(SED) shifts from the optical to soft X-rays as the radius of
the pseudo photosphere shrinks \citep{gallagher78}. Observations
of novae in soft X-rays therefore
generally yield no detections until the photosphere recedes to
the regions within the outflow that are hot enough to produce
X-rays. For some novae this has been observed, and the
X-ray spectra during this phase resemble the class of
Super Soft X-ray Binary Sources \citep[SSS,][]{kahab}.
This phase is therefore called the SSS phase.

Observational evidence (from optical observations) and
theoretical calculations indicate two abundance classes of
novae, those with overabundance of carbon and oxygen (CO
novae) and those with overabundance of oxygen and neon
(ONe novae; see, e.g., \citealt{abunovae94,JH98}). Since
the pressure on the white dwarf surface is not high enough
for the production of C, O, or Ne during the nova outburst,
these abundance classes reflect the composition of the WD.
This indicates that core material is dredged-up into the
accreted material and the gases are mixed before being
ejected into space \citep{S98,G98}. In addition to
dredged-up WD material, the ashes of CNO burning during
the outburst have frequently been observed
\citep{abunovae94,JH98}. The composition of the ejected
material is thus highly non-solar.

RS\,Oph is a Recurrent Symbiotic Nova, which erupts about
every 20 years. The latest outburst occurred on 2006 February
12.83 \citep[=\,day\,0;][]{rsophdiscovery}. The mass donor is
a red giant (M2III), and the expanding ejecta interact
with the pre-existing stellar wind setting up shock systems.
The composition of the red giant was studied by
\cite{pavlenko}, who found that the overall metallicity
does not seem to be significantly different from solar
($[$Fe/H$]=0.0\,\pm\,$0.5), C is underabundant
($[$C$]=-0.4$), and N overabundant ($[$N$]=+0.9$).
UV spectra taken with IUE during the 1985 outburst provided
evidence that N was overabundant \citep{shore96}. Lines of
C were observed, but no detailed abundance analyses were
carried out by \cite{shore96}.
\cite{contini95} determined an N/C abundance ratio of 100
and N/H=10 from optical spectra taken on day 201. From their
absolute abundance of N and Fe, an abundance ratio of
N/Fe=15 relative to solar can be derived. \cite{snijders87}
found N/O=1.1 and C/N=0.16, and they caution that the evolved
secondary can already be C/N depleted.
\cite{contini95} found significant underabundance of O/H
and of Ne/H of $\sim 10$\% solar but high abundance ratios of
Mg/Fe=5.4 and Si/Fe=7.2.

During the first month after outburst, intense hard X-ray
emission, that originated from the shock, was observed
with \swift\ and the Rossi X-ray Timing Explorer (RXTE;
\citealt{bode06,osborne06,sokoloski06}). \swift\ X-Ray Telescope
(XRT) observations carried out between days 3--26 were analyzed
by \cite{bode06} who applied single-temperature MEKAL models to
the X-ray spectra. They determined temperatures and wind column
densities, $N_{\rm W}=N_{\rm H}({\rm total})
-N_{\rm H}({\rm interstellar})$. The interstellar value of
$N_{\rm H}({\rm interstellar})=2.4\times 10^{21}$\,cm$^{-2}$
has been determined from H\,{\sc i}
21\,cm measurements \citep{hje86}. This value is consistent with
the visual extinction ($E(B-V) = 0.73\,\pm\,0.1$) determined from
IUE observations in 1985 \citep{sni87}. \cite{bode06}
converted the temperatures found from the MEKAL models into
derived shock velocities $v_{\rm s}$, assuming
that the X-rays were produced in the blast wave driven
into the circumstellar material following the outburst.
Before day $\sim 6$ after outburst they found a
power-law decay $t^{-\alpha}$ with an
approximate index $\alpha=0.6,\ 0.5$, and $1.5$ for $v_{\rm s}$,
$N_{\rm W}$, and the flux (unabsorbed, i.e., corrected for
interstellar absorption), respectively. These results
compare well with model predictions of the RS\,Oph system
presented by O'Brien et al. (\citeyear{obrien92} - see also
\citealt{bodekahn85}). According to these models, the
evolution can be divided into three phases. The first phase (I),
where the ejecta are still important in supplying energy to the
shocked stellar wind of the red giant, lasts only a few days.
The second phase (II) commences when the blast wave is being
driven into the stellar wind and is effectively adiabatic. This
phase is expected to last
until the shocked material is well cooled by radiation
(phase III). The physics behind these phases of evolution,
together with the density distribution in the wind,
determine the evolution of temperature with the corresponding
velocity of the shock, unabsorbed fluxes, and the
absorbing column of the wind \citep{vaytet07}.

\cite{sokoloski06} analyzed X-ray data taken between days 3--21
with RXTE, and from thermal bremsstrahlung models found
that the temperature decreased with time $t$ as $t^{-2/3}$. They
concluded that the speed
of the blast wave produced in the nova explosion decreased with
$t^{-1/3}$. However, the RXTE data with their low sensitivity
at low energies did not favor the measurement of the wind
column density $N_{\rm W}$.

A \chandra\ High Energy Transmission Grating Spectrograph snapshot
of the blast wave obtained at the end of day 13 and analyzed
by \cite{drake07} shows asymmetric emission lines sculpted by
differential absorption in the circumstellar medium and explosion
ejecta. \cite{drake07} found the lines to be more sharply peaked
than expected for a spherically-symmetric explosion and concluded
that the blast wave was collimated in the direction perpendicular
to the line of sight, as also suggested by contemporaneous radio
interferometry \citep{obrien06}.

 The SSS phase was observed after day $\sim 30$ and ended before
day $\sim 100$ after outburst \citep{osborne06}. Three
high-resolution X-ray spectra were taken during this phase
which are described by \cite{ness_rsoph}. The SSS emission
longward of $\sim 12$\,\AA\ ($E>1$\,keV) outshines any
emission produced by the shock at these wavelengths,
however, all emission shortward of 12\,\AA\ originates
exclusively from the shock \citep{ness_keele,ness_rsoph}.
We note that
\cite{bode_keele} show tentative evidence for emission
between 6-12\,\AA\ that may reflect the evolution of the SSS.
The SSS spectra analyzed by \cite{ness_rsoph} contain
emission lines on top of the bright SSS continuum which,
combined with blue-shifted absorption lines, were first
attributed to P Cygni profiles \citep{rsoph_iau1}, but may
also originate from the shock \cite{ness_rsoph}.

An analysis of all X-ray grating spectra was presented by
\cite{nelson07}. They discovered a soft X-ray flare in week
4 of the evolution in which a new system of low-energy
emission lines appeared. With their identifications of the
emission lines, they derived velocities of
$8,000-10,000$\,km\,s$^{-1}$ which is
consistent with the escape velocity of the WD, and the
new component may thus represent the outflow.
From preliminary atmosphere models they also determined
the abundance ratio of Carbon to Nitrogen of 0.001
solar. This is a factor 10 lower than C/N abundance
measurements by \cite{contini95} and a factor 100 lower
than \cite{snijders87}. From He-like line flux ratios they
confirm that the shock plasma is collisionally dominated.
They measured line shifts and line widths and found that
the magnitude of the velocity shift increases for lower
ionization states and longer wavelengths. In addition,
as the wavelength increased, so did the broadening of
the lines. They discuss bow shocks as a possible origin
for the line emission seen in RS Oph.
 From multi-temperature plasma modelling of the
early X-ray spectra, \cite{nelson07} needed four
temperature components. While they found reasonably good
reproduction of the \chandra\ spectrum, the same model
was in poor agreement with the simultaneous \xmm\ spectrum.
Their model underpredicts lines of O and N, and they
concluded that these elements are overabundant, and that
the lines originated in the ejecta.

The structure of this paper is as follows: In \S\ref{obssect}
we present 12 X-ray grating observations
taken between days 13.8 and 239.2 after outburst, focusing
only on the emission produced by the shock.
We measure emission line fluxes and
line ratios in \S\ref{lines}, and in \S\ref{anal} we
present supporting models. We compute
multi-temperature spectral models with the fitting program
{\sc xspec} (\S\ref{xspecsect}) and reconstruct a continuous
temperature distribution based on a few selected emission
lines (\S\ref{lfluxes}), yielding
the elemental abundances. In \S\ref{cmpmodels} we
compare the results of these two model approaches.
We dedicate a separate section (\S\ref{syserr}) to
the discussion of systematic uncertainties, as all
given error estimates are only statistical uncertainties.
In \S\ref{disc} we discuss our results and
summarize our conclusions in \S\ref{concl}.

\section{Observations}
\label{obssect}

\begin{figure*}[!ht]
\resizebox{\hsize}{!}{\includegraphics{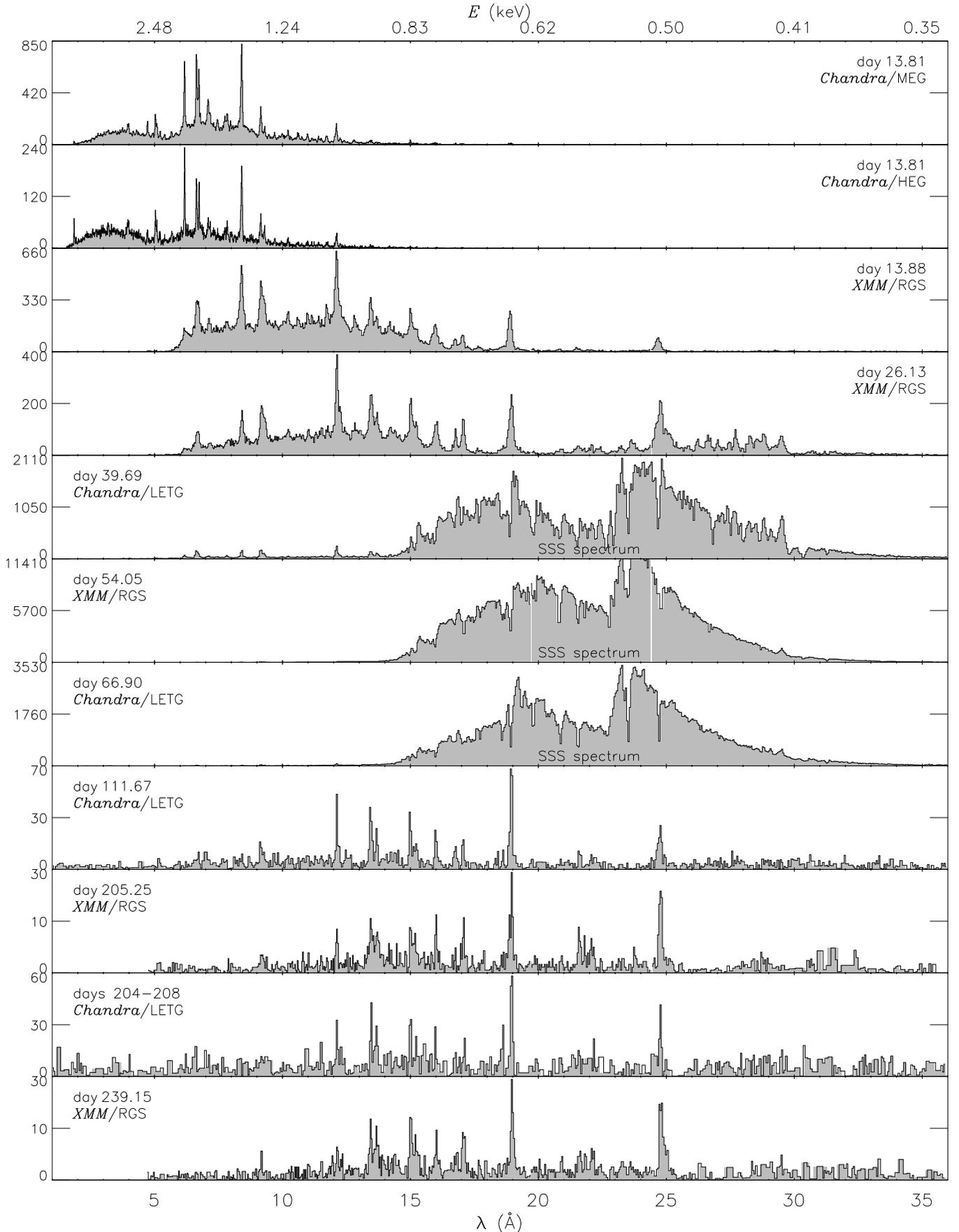}} 
\caption{\label{allspecs}11 X-ray grating spectra 
extracted from 12 \chandra\ and \xmm\ observations
taken on the dates and with the instruments given in
the legends. Plotted are the raw spectra in counts
per bin per individual exposure time (see Table~\ref{tab1}).
Bin sizes are 0.005\,\AA, 0.0025\,\AA, and 0.01\,\AA\ for
MEG, HEG, and RGS1, RGS2, and LETG, respectively. Four
\chandra\ spectra taken between days 204 and 208 after
outburst are combined. RGS1 and RGS2 spectra are combined.}
\end{figure*}

 In this paper we analyze five X-ray grating spectra taken with
\chandra\ and five with \xmm. We use the High- and Low Energy
Transmission Grating spectrometers (HETG and LETG, respectively) aboard
\chandra\ and the Reflection Grating Spectrometers (RGS1 and
RGS2) aboard \xmm\ to obtain data between 1\,\AA\ and 40\,\AA.
In Table~\ref{tab1} we list the start- and stop times,
the corresponding days after outburst, the mission and
instrumental setup, ObsIDs, and net exposure times for each
observation. We have extracted the spectra in the same way as
described by \cite{ness_rsoph} using the standard
tools provided by the mission-specific software packages SAS
(Science Analsis Software, version 7.0) and CIAO (Chandra
Interactive Analysis of Observations, version 3.3.0.1).
While pile up in the zero-th order of the \chandra\ HETGS
observation may lead to problems in centroiding the extraction
regions for the dispersed spectra \citep{nelson07}, we are
confident that the standard centroiding is accurate enough.
For example, the wavelengths of strong lines from the two
opposite dispersion orders agree well with each other.

 We have also extracted spectra from the \xmm\ European Photon
Imaging Camera (EPIC), concentrating on the observations
recorded with the Metal Oxide Semi-conductor chips (MOS1).
We have used standard SAS routines for the extraction of spectra
and have corrected for pile up using annular extraction regions
that avoid extracting photons from the innermost regions
of the point spread function (PSF) following the instructions
provided by the SAS software. We only need the MOS1 data
for ObsID 0410180201 (26.1 days after outburst) in
\S\ref{xspecsect}, for which an inner radius of 300 pix
($15\arcsec$) has to be excluded to avoid significant pile up.

\begin{table}[!ht]
\begin{flushleft}
\renewcommand{\arraystretch}{1.1}
\caption{\label{tab1}Grating observations of RS Oph}
\begin{tabular}{ccclll}
Date & Day$^{a}$& Mission & Grating & ObsID & exp. time\\
start--stop & & & /detector & & (net; ks)\\ 
\hline

\ Febr. 26, 15:20 & 13.81 & \chandra & HETG & 7280 & 9.9\\
--Febr. 26, 18:46 & 13.88 & & /ACIS & &\\
\ Febr. 26, 17:09 & 13.88 & {\it XMM} & \multicolumn{2}{l}{RGS1\ \ 0410180101} & 23.8\\
--Febr. 26, 23:48 & 14.2 & &\multicolumn{2}{l}{RGS2} & 23.8\\
\ March 10, 23:04 & 26.1 & {\it XMM} & \multicolumn{2}{l}{RGS1\ \ 0410180201} & 11.7\\
--March 11, 02:21 & 26.3 & &\multicolumn{2}{l}{RGS2} & 11.7\\

\ March 24, 12:25 & 39.7 & \chandra & LETG & 7296 & 10.0\\
--March 24, 15:38 & 39.8 & & /HRC & &\\
\ April 07, 21:05 & 54.0 & {\it XMM} & \multicolumn{2}{l}{RGS1\ \ 0410180301} & 9.8\\
--April 08, 02:20 & 54.3 & &\multicolumn{2}{l}{RGS2} & 18.6\\
\ April 20, 17:24 & 66.9 & \chandra & LETG & 7297 & 6.5\\
--April 20, 20:28 & 67.0 & & /HRC & &\\

\ June 04, 12:06 & 111.7 & \chandra & LETG & 7298 & 19.9\\
--June 04, 18:08 & 111.9 & & /HRC & &\\

\ Sept. 06, 01:59 & 205.3 & {\it XMM} & \multicolumn{2}{l}{RGS1\ \ 0410180401} & 30.2\\
--Sept. 06, 17:30 & 205.9 & &\multicolumn{2}{l}{RGS2} & 30.2\\

\ Sept. 04, 10:43 & 203.6 & \chandra & LETG & 7390 & 39.6\\
--Sept. 04, 22:26 & 204.1 & & /HRC & &\\
\ Sept. 07, 02:37 & 206.3 & \chandra & LETG & 7389 & 39.8\\
--Sept. 07, 14:29 & 206.8 & & /HRC & &\\
\ Sept. 08, 17:58 & 207.9 & \chandra & LETG & 7403 & 17.9\\
--Sept. 08, 23:36 & 208.2 & & /HRC & &\\

\ Oct. 09, 23:38 & 239.2 & {\it XMM} & \multicolumn{2}{l}{RGS1\ \ 0410180501} & 48.7\\
--Oct. 10, 13:18 & 239.7 & &\multicolumn{2}{l}{RGS2} & 48.7\\

\hline
\end{tabular}

$^{a}$after outburst (2006, Feb. 12.83)
\renewcommand{\arraystretch}{1}
\end{flushleft}
\end{table}

\subsection{Description of spectra}

An overview of all grating spectra is presented in
Fig.~\ref{allspecs}
with the instrument and day after outburst indicated in the
legends of each panel. All spectra taken on days 13.8 and
26.1 after outburst (top three panels) are characterized by a
hard, broad continuum spectrum with additional strong emission
lines \citep{rsoph_iau1,drake07}. The count rate on day 26.1
is significantly lower than that on day  13.8, and the shape of
the continuum
is different. On day 26.1 a new component is observed longward
of $\sim 20$\,\AA\ \citep{nelson07} that could be associated with
the SSS spectrum that was clearly detected three days later with
\swift\ \citep{osborne06}. However, the spectral shape of this
new component on day 26.1 is quite different from the spectra
observed on days 39.7, 54.0, and 66.9 (next three panels).
These spectra are
dominated by the SSS spectrum \citep{ness_rsoph} between
14\,\AA\ and 37\,\AA, while the emission from the shock
dominates shortward of $\sim 15$\,\AA\ \citep{ness_keele}. After
day $\sim 100$, the SSS spectrum has disappeared, and those
spectra display emission lines with a weak continuum.
The short-wavelength lines are only seen in the early spectra
while those between 12\,\AA\ and 25\,\AA\ can be seen in
all spectra, however, with different relative strengths.

\begin{figure*}[!ht]
\resizebox{\hsize}{!}{\includegraphics{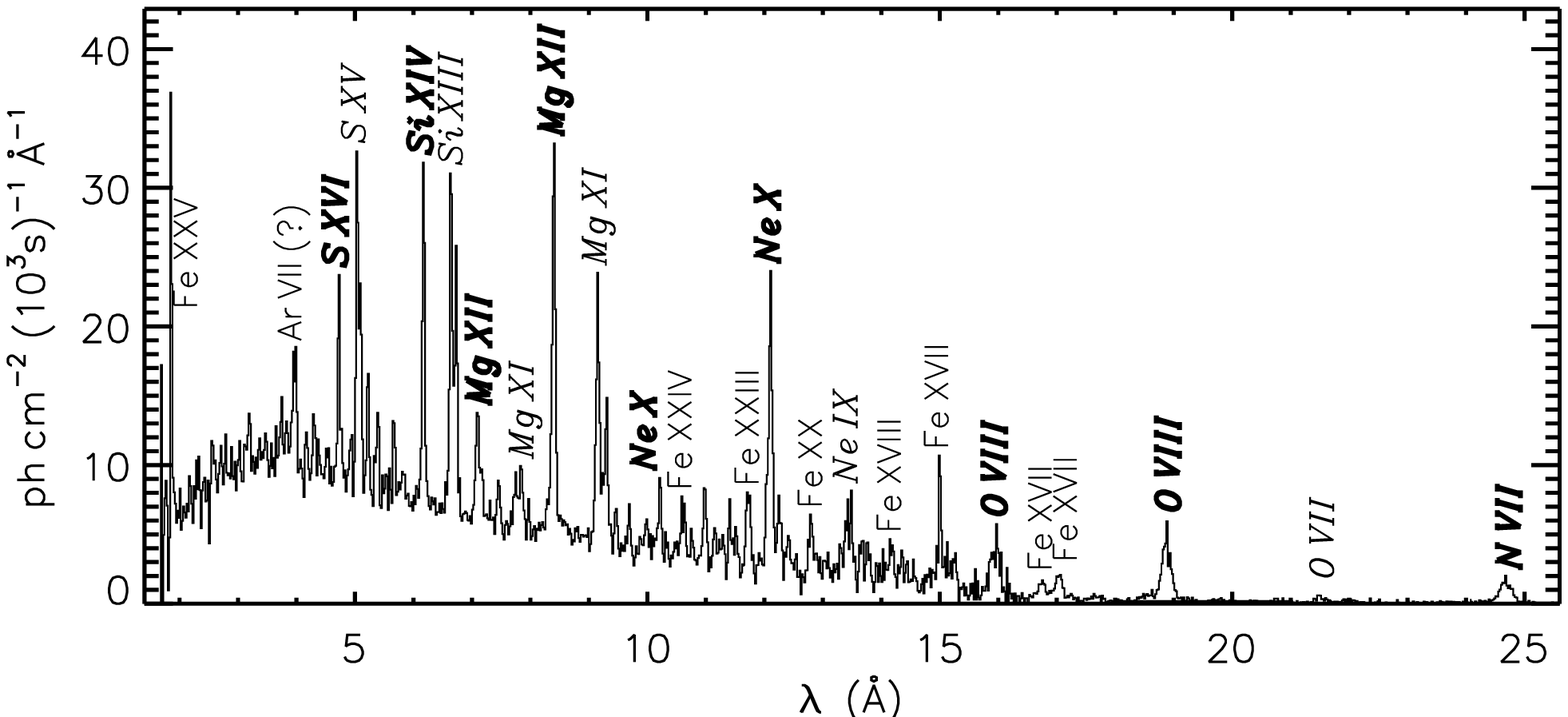}}
\caption{\label{day13.8}X-ray spectrum of RS\,Oph on
day 13.8, in photon flux units, taken with \chandra/MEG
shortward of 16.5\,\AA\ and with \xmm/RGS longwards.
We label H-like and He-like lines in italic font with
the H-like lines in bold-face. Other lines are
labeled with roman font. The high ionization stages indicate
temperatures up to $10^8$\,K while the additional presence
of low-ionization stages show that the plasma is not
isothermal.
}
\end{figure*}

In Fig.~\ref{day13.8} we show the X-ray spectrum taken
on day 13.8. For this plot we have converted the number of
counts in each spectral bin to photon fluxes, simply dividing
the number of counts by the effective areas extracted for
each spectral bin from the instrument calibration. With
grating spectra such a conversion is sufficiently accurate
because of the precise
placement of the recorded photons into the spectral grid.
In contrast to low-resolution X-ray spectra taken with
CCDs, the photon redistribution matrix of grating spectra
is nearly diagonal. Below 16.5\,\AA\ we show the
\chandra/MEG spectrum, and above this wavelength,
where the MEG has extremely low sensitivity, the
combined \xmm/RGS spectra are shown. The strongest lines
seen in
the spectrum originate from H-like and He-like ions of
S\,{\sc xvi} and S\,{\sc xv} (4.73 and 5.04\,\AA),
Si\,{\sc xiv} and Si\,{\sc xiii} (6.18 and 6.65\,\AA),
Mg\,{\sc xii} and Mg\,{\sc xi} (8.42 and 9.2\,\AA),
Ne\,{\sc x} and Ne\,{\sc ix} (12.1 and 13.5\,\AA),
O\,{\sc viii} and O\,{\sc vii} (18.97 and 21.6\,\AA),
and N\,{\sc vii} (24.78\,\AA). Also some of the
3p-1s lines are detected, e.g., Mg\,{\sc xii} at
7.11\,\AA, Mg\,{\sc xi} at 7.85\,\AA, and O\,{\sc viii}
at 16\,\AA. The H-like and He-like lines of elements
with higher nuclear charge arise at shorter wavelengths,
and strong lines at short wavelength indicate
high temperatures. Several Fe lines are present,
e.g., Fe\,{\sc xxv} (1.85+1.86+1.87\,\AA) and
Fe\,{\sc xxiv} at 10.62\,\AA\ as well as
low-ionization lines of Fe\,{\sc xvii} at 15.01\,\AA\
and 12.26\,\AA. These lines cannot be formed in the
same region of the plasma and it is thus not isothermal.

\begin{figure*}[!ht]
\resizebox{\hsize}{!}{\includegraphics{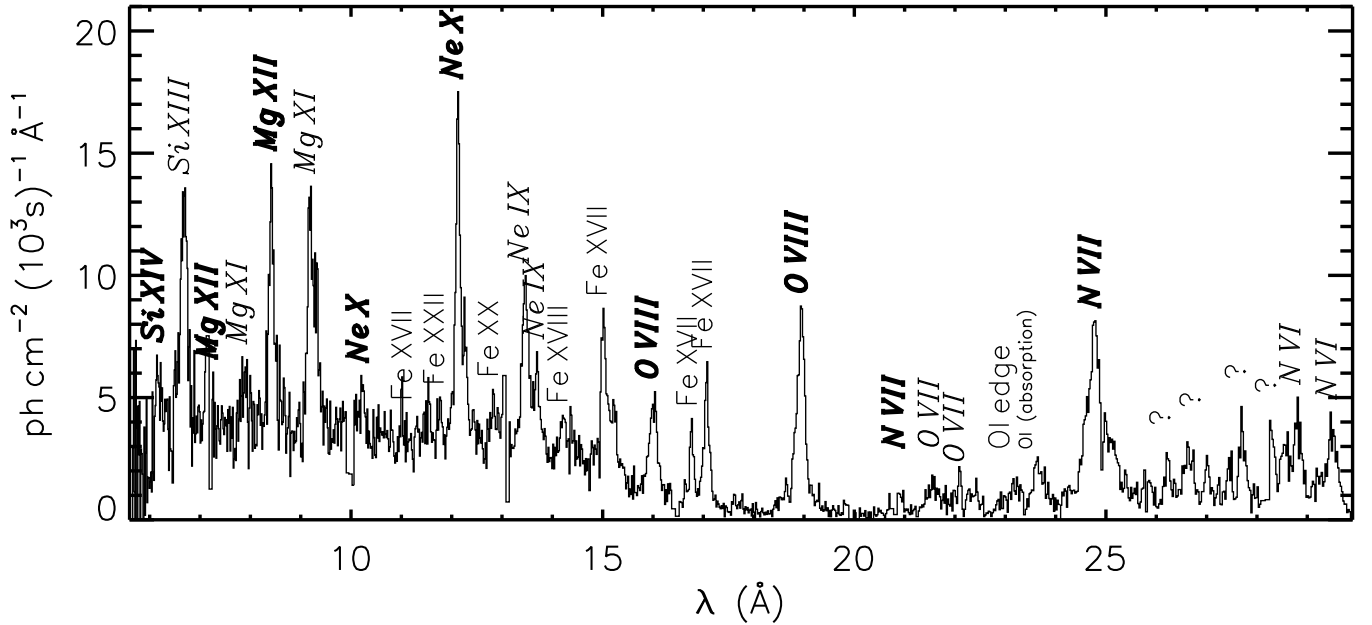}}
\caption{\label{day26.1}\xmm/RGS1 and RGS2 spectra (combined)
taken on day 26.1 and converted to photon flux units. The
strongest emission lines are labelled as explained in
Fig.~\ref{day13.8}.
}
\end{figure*}

 In Fig.~\ref{day26.1} we show the combined \xmm/RGS spectra
taken on day 26.1, in the same units as in Fig.~\ref{day13.8}
for direct comparison. While on day 13.8 the strongest lines
are formed at wavelengths shortward of 10\,\AA, the Ne\,{\sc x}
line at 12.1\,\AA\ is now the strongest line. This could
mean that the temperature and/or the neutral
hydrogen column density have decreased. The relative
line strengths of H-like to He-like lines are significantly
lower for all elements (see, e.g., Mg\,{\sc xii} to
Mg\,{\sc xi}). This is clearly a temperature effect, and
the plasma is cooling. Longwards of 25\,\AA\ a new component
can be seen. The fact that only three days later the SSS
spectrum was observed with \swift\ \citep{osborne06}
suggests that this emission represents the onset of the SSS
phase \citep[e.g.,][]{bode06,nelson07}. However,
while the SSS spectra observed on day 39.7 range from
$\sim 15-30$\,\AA\ (Fig.~\ref{allspecs}), the RGS spectra
shown in Fig.~\ref{day26.1} show only excess emission
longward of $\sim 20$\,\AA\ (see also Fig.~\ref{mepic}). At
23.5\,\AA\ a deep absorption edge from O\,{\sc i} has been
found in the SSS spectra of RS\,Oph by \cite{ness_rsoph}
(see also bottom panel of Fig.~\ref{mepic}).
The hard portion of an early faint SSS spectrum might be
entirely absorbed by circumstellar neutral oxygen in the
line of sight, while the shock-induced emission may originate
from further outside, thus traversing through less absorbing
material. Also, in the standard picture of
nova evolution, the peak of the SED is
expected to shift from long wavelengths to short wavelengths
while the radius of the photosphere recedes to successively
hotter layers, and the observed emission would be consistent
with this picture. However, the spectrum has more
characteristics of an emission line spectrum (see
Fig.~\ref{day26.1} and \citealt{nelson07}), but only the
lines at 24.79\,\AA\ and 28.78+29.1+29.54\,\AA\ can be
identified as N\,{\sc vii} and as the N\,{\sc vi} He-like
triplet lines, respectively. In between the N\,{\sc vii}
and N\,{\sc vi} lines no strong lines are listed
in any of the atomic databases. The strongest emission
line in this range is observed as a narrow line at
27.7\,\AA\ (FWHM 0.08\,\AA) with a line flux of
$(2.7\,\pm\,0.4)\times10^{-13}$\,erg\,cm$^{-2}$\,s$^{-1}$.
The only possible identifications would be
Ar\,{\sc xiv} (27.64\,\AA\ and 27.46\,\AA) or
Ca\,{\sc xiv} (27.77\,\AA). Both appear rather unlikely
identifications, as no Ar lines are detected in any
of the other spectra, and for Ca\,{\sc xiv}, stronger
lines are expected at 24.03\,\AA, 24.09\,\AA, and
24.13\,\AA, but are not detected. A remarkable aspect is that
the 27.7-\AA\ line is so narrow while the N\,{\sc vii} line
shows an extremely broad profile (see \S\ref{lprofiles}).
Another unidentified line is measured at 23.6\,\AA, but we
experience the same difficulties in finding an identification.
This could be residual continuum emission if the absorption
feature at 23.5\,\AA\ is interpreted as interstellar
O\,{\sc i}. \cite{nelson07} suggested that some of these
lines are blue-shifted N\,{\sc vi} and C\,{\sc vi} lines,
but this requires extremely high velocities and is in
contradiction to the non-detection of the C\,{\sc vi}
Ly$\alpha$ line and the low C abundance reported in the same
paper. In any case, this component is likely not part of
the shock systems, and the discussion of it is beyond
the scope of this paper. While the O\,{\sc viii} and
O\,{\sc vii} lines might be part of the shock, we treat
the interpretation of these lines with care. We will also include
the N\,{\sc vii} line in our analyses as if it were formed
in the shock, and any inconsistencies based on this line
can be understood as supporting evidence that this component
is unrelated to the shock emission. For more details of this
component we also refer to \cite{nelson07}.

\begin{figure*}[!ht]
\resizebox{\hsize}{!}{\includegraphics{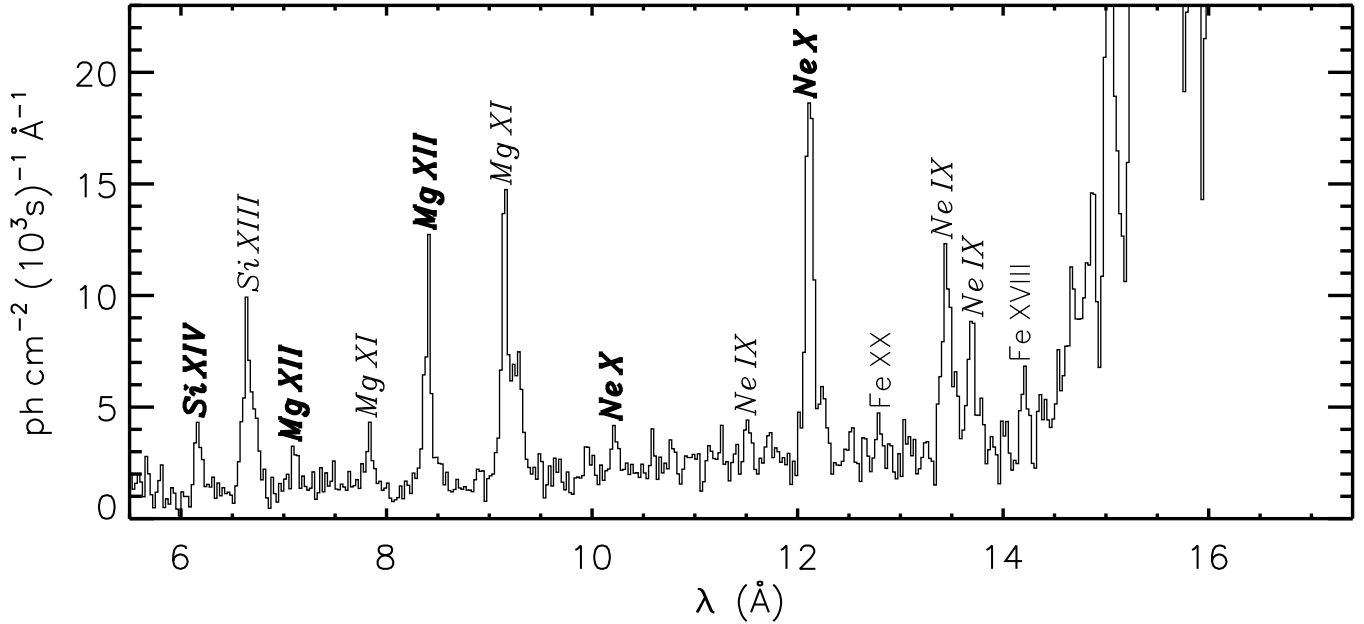}}
\caption{\label{day39.7}Same as Fig.~\ref{day26.1} for
the \chandra/LETG spectrum taken on day 39.7. Longward of
$\sim 14.5$\,\AA\ the SSS spectrum dominates.
}
\end{figure*}

In Fig.~\ref{day39.7} we show the photon flux spectrum taken
with \chandra\ LETGS on day 39.7. The SSS spectrum dominates
all emission longward of $\sim 14.5$\,\AA, and we only show the
wavelength range relevant for this paper. The ratio of H-like
to He-like lines is lower than in the earlier spectra,
indicating that the temperature has continued to decrease.
Since the lines are formed shortwards of the high-energy (Wien)
tail of the SSS
spectrum (14.5\,\AA$\approx 0.86$\,keV), they are not
affected by photoexcitations and originate exclusively from
the shock.

\begin{figure*}[!ht]
\resizebox{\hsize}{!}{\includegraphics{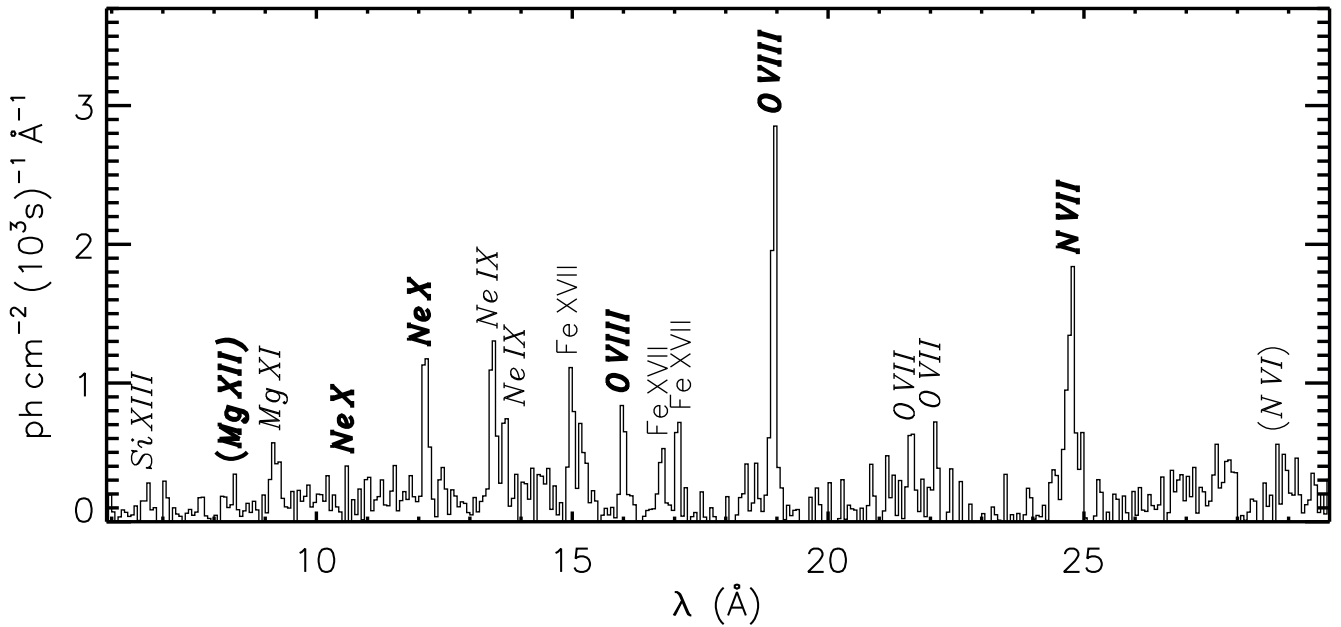}}
\caption{\label{day111.7}Same as Fig.~\ref{day26.1} for
the \chandra/LETG spectrum taken on day 111.7. The SSS spectrum
has disappeared and emission lines longward of $\sim 14.5$\,\AA\
can be observed.}
\end{figure*}

In Fig.~\ref{day111.7} we show one of the spectra taken
after the SSS had turned off. All emission lines are significantly
weaker and the ratio of H-like and He-like lines is again
lower than in the previous observation. All short-wavelength
lines are extremely weak or are not detected.

\begin{figure}[!ht]
\resizebox{\hsize}{!}{\includegraphics{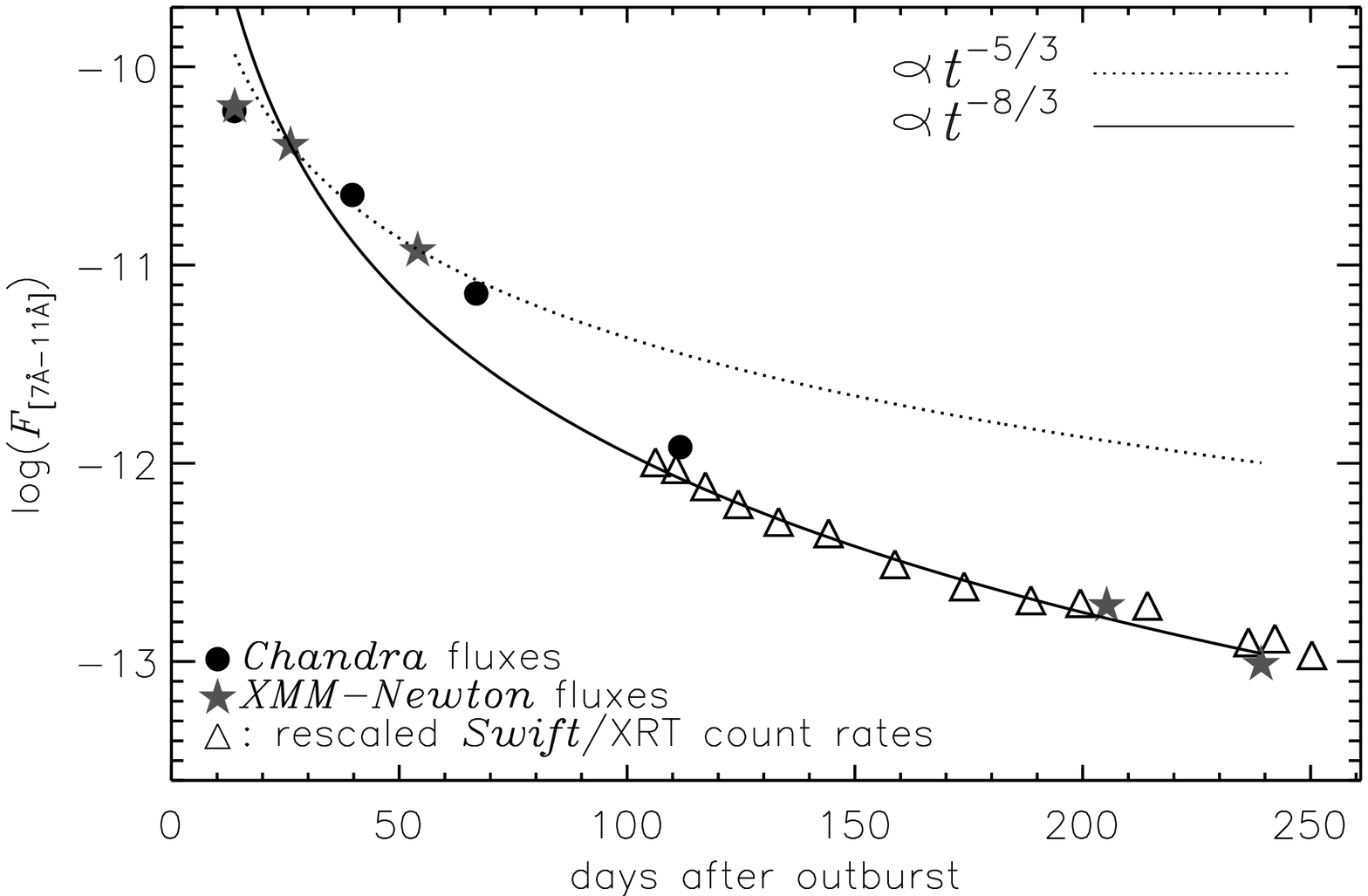}}
\caption{\label{evol}X-ray fluxes measured at Earth
(cgs units, integrated over 7--11\,\AA;
1.1--1.8\,keV) as a function of time. The triangles
mark \swift/XRT count rates (0.3-10\,keV), rescaled by
$3\times 10^{-12}$\,[erg\,cm$^{-2}$\,s$^{-1}$\,cps$^{-1}$].
The statistical errors are smaller than the plot
symbols. For systematic errors see \S\ref{syserr}.
}
\end{figure}

Next we integrate the photon flux spectra over the range
7--11\,\AA\ (1.1--1.8\,keV) in order to obtain X-ray fluxes.
We do not correct for absorption, thus yielding fluxes at Earth.
Since the fluxes are extracted from above 1\,keV, the effects
from absorption are small, and particularly the relative
evolution of the absorbed and non-absorbed fluxes is the same.
The wavelength range over which the fluxes are integrated
is a compromise between collecting as much information
as possible from the observations before day 39.7 and after day
66.9 while excluding as much as possible of the emission from
the SSS on days 39.7, 54.0, and 66.9. The results are illustrated
as a function of time in Fig.~\ref{evol}. For comparison we
include rescaled \swift/XRT count rates (0.25-10\,keV) taken
after day 106. At this late stage of the evolution, the spectral
shape hardly changes (see Table~\ref{xspec}),
yielding a direct correlation between X-ray flux and count rate.
With the assumption of no spectral changes between days 106 and
250, we can also use the count rates integrated over the full
\swift\ XRT band pass, as additional emission in the larger
wavelength range also scales directly with the count rate.
Since we are not interested in the absolute flux from the
\swift\ observations, we chose a scaling factor of
$3\times10^{-12}$\,erg\,cm$^{-2}$\,s$^{-1}$\,cps$^{-1}$
to yield the same values as the grating fluxes for
days 111.7-239.2. The rescaled XRT count rate follows the
same trend as the fluxes obtained from the grating spectra.
 We include two power-law curves,
and the early evolution evolves more like $t^{-5/3}$, while
the later evolution (after day 100) clearly follows a
$t^{-8/3}$ trend. We observe the same behavior if we use
a larger wavelength range and exclude the observations
between days 39.7 and 66.9.

\section{Measurement of emission lines}
\label{lines}

\subsection{Line shifts and profiles}
\label{lprofiles}

 In order to determine velocities from the emission lines we have
measured wavelengths and line widths in excess of the instrumental
line broadening function for a number of strong lines with
well-identified rest wavelengths, $\lambda_0$. For the narrow
wavelength range around the lines we have accounted for continuum
emission by defining a constant local offset on top of the
instrumental background that can be treated as an 'uninteresting'
free parameter. For each line $j$ we have used a normalized Gaussian
profile with wavelength $\lambda_j$
and line width $\sigma_j$ and folded this profile through the
instrumental response using the IDL tool {\tt scrmf} provided by
the PINTofAle package \citep{pintofale} before comparing with the
measured count spectra. We have determined the statistical
measurement uncertainties for
$\lambda_j$ and $\sigma_j$ from the $2\times 2$ Hesse matrix as
defined in Eq.~\ref{hesse} of the appendix section, which is
based on an approach proposed by \cite{strong85}.
Systematic uncertainties are
difficult to assess and are not included in our error estimates
(for details see \S\ref{syserr}). Those can arise from
fluctuations in the underlying continuum and line blends.
While the former has a stronger effect on weak lines, the latter
can affect any line. For this reason we chose lines for which no
strong nearby lines are known to arise. We have
iterated $\lambda_j$ and $\sigma_j$, and in each iteration
step we have adjusted the normalization utilizing the fixed point
iteration scheme described by \cite{newi02}. The normalization
factor can be converted to line fluxes (see \S\ref{lfluxmeas}).

The results are listed in Table~\ref{vlines}. The line shifts
$\lambda_j-\lambda_0$ and Gaussian line widths $\sigma_j$
(both measured in m\AA$=10^{-3}$\AA) are converted
to corresponding Doppler velocities using the rest wavelengths
$\lambda_0$ listed in the first column. The measurement
uncertainties of line shifts and widths are correlated
uncertainties, and account for the uncertainties in
the respective other values. The Ne\,{\sc x} line at 10.23\,\AA\
is relatively weak in all observations, and the results from
this line may be less certain due to additional systematic
uncertainties from fluctuations in the underlying continuum.
The Fe\,{\sc xvii}
line could be blended with the weak O\,{\sc viii} 1s-4p
($\lambda_0=15.18$\,\AA) line,
and the accuracy of the results from this line might suffer from
line blending. All lines measured from observations taken after
day 26.1 are weaker, and the uncertainties on the results from
these observations have to be increased by at least 20\% due to
fluctuations in the continuum.

\begin{figure}[!ht]
\resizebox{\hsize}{!}{\includegraphics{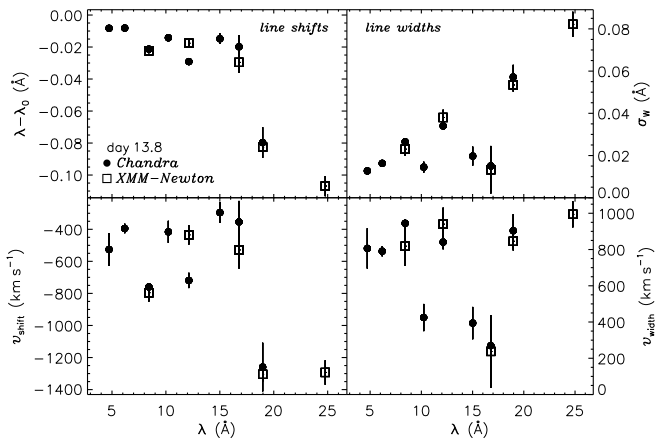}}
\caption{\label{v13}Measurement of line shifts (left panels)
and line widths (right panels) for the \chandra\ HETGS
(bullets) and \xmm\ (open boxes) observations taken on
day 13.8 with the conversion to velocities, if interpreted
as Doppler velocities (bottom panels). The error bars
are statistical uncertainties only.
For systematic errors see \S\ref{syserr}.}
\end{figure}

In Fig.~\ref{v13} we illustrate the measured line
shifts (top left) and widths (top right) and the corresponding
velocities (respective bottom panels) for the observations taken
on day 13.8. All lines are significantly blue-shifted, the
short-wavelength lines by $200-800$\,km\,s$^{-1}$ and the
lines of O\,{\sc viii}, O\,{\sc vii}, and N\,{\sc vii}
at longer wavelengths by more than $1200$\,km\,s$^{-1}$.
\cite{nelson07} found similar values and concluded that
there was a trend of increased velocities with wavelength
and thus with formation temperature. However, with a
different set of lines we come to a different conclusion.
First, we have not used the unresolved He triplet lines
of Mg\,{\sc xi} and Si\,{\sc xiii} to avoid additional
systematic uncertainties from line blends (see \S\ref{syserr}).
Then,
we have included the Ne\,{\sc x} and Fe\,{\sc xvii} lines
that lie in between the Mg lines and the O\,{\sc viii} line.
Although we caution that these lines may suffer from
additional systematic uncertainties, there
seems to be more of an abrupt change rather than a
systematic trend with these additional lines included.
Interestingly, the lines
with larger blue shifts originate only from oxygen and nitrogen.
\cite{drake07} investigated the possibility that the line
profiles are dominated by complex absorption patterns in their
red wings, leading to apparent blue shifts. In that case the
column density in the respective line is a stronger driver for
line shifts than the temperature, and oxygen and nitrogen
might exhibit deeper column densities than other elements,
possibly owing to higher elemental abundances.

The line widths are all about
$800-1000$\,km\,s$^{-1}$, with the exceptions of
Ne\,{\sc x} (10.23\,\AA) and Fe\,{\sc xvii}
(15.01\,\AA\ and 16.78\,\AA) which are narrower
(bottom right panel of Fig.~\ref{v13}). Since fluctuations
in the continuum and line blending cannot lead to narrower
lines, it is
not clear to us why these particular lines are narrower,
but we cannot confirm a trend with long-wavelength
lines being broader as reported by \cite{nelson07}.
Nelson et al. seem not to have accounted for
the instrumental line broadening when computing
line widths, but the instrumental line broadening
is only $\sim 0.01$\,\AA\ for the MEG and
$\sim 0.03$\,\AA\ in the RGS. The instrumental
line profile is roughly Gaussian, and since the
convolution of two Gaussians is again a Gaussian,
the resulting line width is dominated by the
broader line, and in the case of most lines, the
instrumental line broadening can be neglected.

After day 13.8, all line shifts except those for the
O\,{\sc viii} and N\,{\sc vii} lines fluctuate around
the same value of $\sim 500-800$\,km\,s$^{-1}$ (see
Table~\ref{vlines}). The O and N lines that show
extreme blue-shifts on day 13.8 (Fig.~\ref{v13}) have
values consistent with other lines in all later
observations. If these lines are shaped by absorption in
their red wings as proposed by \cite{drake07}, then the
column densities have decreased from
day 13.8 to day 261. The N\,{\sc vii} line on day 26.1
shows an extreme value (also in line width, see bottom
panel of Fig.~\ref{v13}), and belongs to the new component discovered by
\cite{nelson07}; however, the velocity measured
from the shift of this line does not agree with their
value of 8,000-10,000\,km\,s$^{-1}$ derived from the
lines between 25--30\,\AA. While we mark these lines as
unidentified in Fig.~\ref{day26.1}, \cite{nelson07}
discuss possible identifications as highly blue-shifted
N\,{\sc vi} and C\,{\sc vi} lines.

 The line widths slowly decrease with time. The N\,{\sc
vii} line at 24.78\,\AA\ is extremely broad on day 26.1.
At this time of the evolution, this line is part of the
new component reported by \cite{nelson07} with a set
of unidentified lines. There is thus a reasonable chance
that this line is a blend, making this anomalous velocity
questionable. Since shock
velocities derived from the temperatures from the spectral
models discussed in \S\ref{xspecsect} represent
the evolution of the expansion velocity, they can
be compared with these values.
The shock velocities derived from the hottest
model component are given in the last row of
Table~\ref{vlines}. We discuss the implication of the
comparison in \S\ref{xspecsect}.

\begin{table*}[!ht]
\begin{flushleft}
\renewcommand{\arraystretch}{1.1}
\caption{\label{vlines}Evolution of line shifts and line widths.}
\begin{tabular}{lcccccccc}
$\lambda_0^{a}$\small{(\AA)}&day 13.81 & day 13.88 & day 26.1 & day 39.7 & day 54.0 & day 66.9 & day 111.7 \\
ID&HETG&RGS&RGS&LETG&RGS&LETGS&LETGS\\
\hline
{\bf 4.73}\ \hfill $\Delta\lambda$\,{\tiny (m\AA)} & $-8.3\,\pm\,1.6$ & -- & -- & -- & -- & -- & --\\
{\bf S\,{\sc xvi}}\hfill $\sigma$\,{\tiny (m\AA)} & $12.7\,\pm\,1.7$ & -- & -- & -- & -- & -- & --\\
\hfill $v_{\rm shift}$\,{\tiny (km\,s$^{-1}$)} & $-526\,\pm\,103$ & -- & -- & -- & -- & -- & --\\
\hfill $v_{\rm width}$\,{\tiny (km\,s$^{-1}$)} & $806\,\pm\,111$ & -- & -- & -- & -- & -- & --\\
\hline
{\bf 6.18}\ \hfill $\Delta\lambda$\,{\tiny (m\AA)} & $-8.1\,\pm\,0.6$ & $-4.8\,\pm\,5.7$ & -- & $-16.1\,\pm\,5.2$ & -- & -- & --\\
{\bf Si\,{\sc xiv}}\hfill $\sigma$\,{\tiny (m\AA)} & $16.3\,\pm\,0.6$ & $<12.1$ & -- & $17.7\,\pm\,8.9$ & -- & -- & --\\
\hfill $v_{\rm shift}$\,{\tiny (km\,s$^{-1}$)} & $-395\,\pm\,29$ & $-233\,\pm\,275$ & -- & $-781\,\pm\,253$ & -- & -- & --\\
\hfill $v_{\rm width}$\,{\tiny (km\,s$^{-1}$)} & $790\,\pm\,27$ & $<586.2$ & -- & $861\,\pm\,433$ & -- & -- & --\\
\hline
{\bf 8.42}\ \hfill $\Delta\lambda$\,{\tiny (m\AA)} & $-21.3\,\pm\,0.7$ & $-22.4\,\pm\,1.5$ & $-6.5\,\pm\,2.9$ & $-25.9\,\pm\,3.0$ & $-17.5\,\pm\,0.7$ & $-16.7\,\pm\,8.8$ & $2.3\,\pm\,18.0$\\
{\bf Mg\,{\sc xii}}\hfill $\sigma$\,{\tiny (m\AA)} & $26.5\,\pm\,0.6$ & $23.0\,\pm\,3.0$ & $20.3\,\pm\,5.3$ & $24.0\,\pm\,5.3$ & $<17.1$ & $<11.0$ & $<27.7$\\
\hfill $v_{\rm shift}$\,{\tiny (km\,s$^{-1}$)} & $-759\,\pm\,24$ & $-796\,\pm\,54$ & $-230\,\pm\,102$ & $-924\,\pm\,107$ & $-623\,\pm\,25$ & $-593\,\pm\,313$ & $82\,\pm\,640$\\
\hfill $v_{\rm width}$\,{\tiny (km\,s$^{-1}$)} & $945\,\pm\,21$ & $818\,\pm\,108$ & $724\,\pm\,189$ & $853\,\pm\,187$ & $<607.8$ & $<390.6$ & $<985.1$\\
\hline
{\bf 10.23}\ \hfill $\Delta\lambda$\,{\tiny (m\AA)} & $-14.2\,\pm\,2.3$ & -- & -- & -- & $-2.2\,\pm\,8.4$ & $-17.9\,\pm\,10.7$ & $-1.3\,\pm\,16.4$\\
{\bf Ne\,{\sc x}}\hfill $\sigma$\,{\tiny (m\AA)} & $14.5\,\pm\,2.6$ & -- & -- & -- & $<11.3$ & $<25.7$ & $<12.1$\\
\hfill $v_{\rm shift}$\,{\tiny (km\,s$^{-1}$)} & $-416\,\pm\,69$ & -- & -- & -- & $-64\,\pm\,246$ & $-525\,\pm\,314$ & $-38\,\pm\,482$\\
\hfill $v_{\rm width}$\,{\tiny (km\,s$^{-1}$)} & $425\,\pm\,75$ & -- & -- & -- & $<331.7$ & $<753.7$ & $<355.1$\\
\hline
{\bf 12.13}\ \hfill $\Delta\lambda$\,{\tiny (m\AA)} & $-29.1\,\pm\,1.9$ & $-17.6\,\pm\,2.4$ & $-6.1\,\pm\,2.5$ & $-19.3\,\pm\,2.6$ & $-3.5\,\pm\,3.3$ & -- & $-6.8\,\pm\,5.6$\\
{\bf Ne\,{\sc x}}\hfill $\sigma$\,{\tiny (m\AA)} & $34.0\,\pm\,1.6$ & $38.0\,\pm\,3.7$ & $24.3\,\pm\,4.4$ & $27.3\,\pm\,4.0$ & $29.7\,\pm\,5.6$ & -- & $<19.0$\\
\hfill $v_{\rm shift}$\,{\tiny (km\,s$^{-1}$)} & $-719\,\pm\,47$ & $-435\,\pm\,60$ & $-151\,\pm\,62$ & $-477\,\pm\,65$ & $-87\,\pm\,81$ & -- & $-168\,\pm\,139$\\
\hfill $v_{\rm width}$\,{\tiny (km\,s$^{-1}$)} & $841\,\pm\,39$ & $939\,\pm\,91$ & $601\,\pm\,110$ & $675\,\pm\,99$ & $733\,\pm\,137$ & -- & $<469.3$\\
\hline
{\bf 15.01}\ \hfill $\Delta\lambda$\,{\tiny (m\AA)} & $-14.8\,\pm\,3.3$ & $12.7\,\pm\,4.8$ & $4.1\,\pm\,3.2$ & -- & -- & -- & $18.3\,\pm\,4.5$\\
{\bf Fe\,{\sc xvii}}\hfill $\sigma$\,{\tiny (m\AA)} & $19.7\,\pm\,4.4$ & $65.5\,\pm\,5.8$ & $30.5\,\pm\,4.4$ & -- & -- & -- & $28.3\,\pm\,3.0$\\
\hfill $v_{\rm shift}$\,{\tiny (km\,s$^{-1}$)} & $-296\,\pm\,65$ & $254\,\pm\,96$ & $82\,\pm\,64$ & -- & -- & -- & $365\,\pm\,91$\\
\hfill $v_{\rm width}$\,{\tiny (km\,s$^{-1}$)} & $394\,\pm\,89$ & $1308\,\pm\,115$ & $609\,\pm\,89$ & -- & -- & -- & $565\,\pm\,60$\\
\hline
{\bf 16.78}\ \hfill $\Delta\lambda$\,{\tiny (m\AA)} & $-19.8\,\pm\,7.3$ & $-29.6\,\pm\,6.5$ & -- & -- & -- & -- & $-38.9\,\pm\,8.5$\\
{\bf Fe\,{\sc xvii}}\hfill $\sigma$\,{\tiny (m\AA)} & $15.1\,\pm\,7.3$ & $13.3\,\pm\,11.2$ & -- & -- & -- & -- & $<14.1$\\
\hfill $v_{\rm shift}$\,{\tiny (km\,s$^{-1}$)} & $-354\,\pm\,131$ & $-529\,\pm\,116$ & -- & -- & -- & -- & $-695\,\pm\,151$\\
\hfill $v_{\rm width}$\,{\tiny (km\,s$^{-1}$)} & $270\,\pm\,130$ & $238\,\pm\,201$ & -- & -- & -- & -- & $<252.6$\\
\hline
{\bf 18.97}\ \hfill $\Delta\lambda$\,{\tiny (m\AA)} & $-79.7\,\pm\,9.6$ & $-82.6\,\pm\,2.8$ & $-38.1\,\pm\,5.2$ & -- & -- & -- & $-26.3\,\pm\,3.0$\\
{\bf O\,{\sc viii}}\hfill $\sigma$\,{\tiny (m\AA)} & $57.2\,\pm\,5.7$ & $53.5\,\pm\,3.1$ & $68.3\,\pm\,3.0$ & -- & -- & -- & $<1.0$\\
\hfill $v_{\rm shift}$\,{\tiny (km\,s$^{-1}$)} & $-1260\,\pm\,151$ & $-1305\,\pm\,44$ & $-602\,\pm\,82$ & -- & -- & -- & $-416\,\pm\,47$\\
\hfill $v_{\rm width}$\,{\tiny (km\,s$^{-1}$)} & $904\,\pm\,91$ & $845\,\pm\,50$ & $1080\,\pm\,48$ & -- & -- & -- & $<16.1$\\
\hline
{\bf 24.78}\ \hfill $\Delta\lambda$\,{\tiny (m\AA)} & -- & $-106.9\,\pm\,6.3$ & $30.0\,\pm\,7.7$ & -- & -- & -- & $-6.9\,\pm\,0.4$\\
{\bf N\,{\sc vii}}\hfill $\sigma$\,{\tiny (m\AA)} & -- & $82.2\,\pm\,5.9$ & $170.3\,\pm\,1.5$ & -- & -- & -- & $<7.9$\\
\hfill $v_{\rm shift}$\,{\tiny (km\,s$^{-1}$)} & -- & $-1293\,\pm\,76$ & $363\,\pm\,93$ & -- & -- & -- & $-83\,\pm\,4$\\
\hfill $v_{\rm width}$\,{\tiny (km\,s$^{-1}$)} & -- & $994\,\pm\,72$ & $2060\,\pm\,18$ & -- & -- & -- & $<95.7$\\
\hline
$v_{\rm shock}(T_1)$\,{\tiny (km\,s$^{-1}$)}$^b$&
\multicolumn{2}{c}{$1899\,\pm\,22$} &  $1284\,\pm\,30$ &  $1269\,\pm\,44$ & -- & -- & $792\,\pm\,37$\\
$v_{\rm shock}(T_2)$\,{\tiny (km\,s$^{-1}$)}$^b$&
\multicolumn{2}{c}{$792\,\pm\,8$} &  $792\,\pm\,8$ &  $714\,\pm\,7$ & -- & $747\,\pm\,60$ &  $548\,\pm\,61$\\

\hline
\end{tabular}

$^a$rest wavelengths\ \ $^b$From k$T_1$ and k$T_2$ in APEC models
\renewcommand{\arraystretch}{1}
\end{flushleft}
\end{table*}

\subsection{Line fluxes}
\label{lfluxmeas}

 We have used our line fitting program Cora \citep{newi02} to measure
line counts which are then converted to line fluxes using the
effective areas extracted from the instrument calibration.
The Cora program applies the likelihood method described in
the appendix and adds a model of line templates to the
instrumental background. In order to measure the line fluxes on
top of the continuum, we have added a constant source background
to the instrumental background before fitting. The continuum
is not expected to change significantly over the narrow
wavelength range considered for line fitting.

 In Table~\ref{elines} we list
the measured line fluxes for the strongest lines of Fe and the
H-like and He-like lines of N, O, Ne, Mg, Si, and S, sorted
by wavelength (for emission line fluxes measured on days
39.7, 66.9, and 54 we refer to \citealt{ness_rsoph}). The
given uncertainties are the statistical uncertainties. Additional
systematic uncertainties arise from the assumed level of the
underlying continuum and line widths (see \S\ref{syserr}). The
fluxes are not corrected for the effects of interstellar or
circumstellar absorption. In the bottom part of Table~\ref{elines}
we list line ratios of H-like to He-like line fluxes, which are
temperature and density indicators.

\begin{figure}[!ht]
\resizebox{\hsize}{!}{\includegraphics{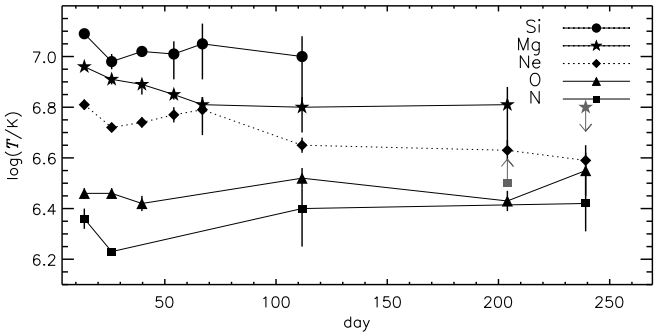}}
\caption{\label{lrats}Evolution of average temperatures
derived from ratios of H-like and He-like lines for the
elements given in the legend. The values are listed in
Table~\ref{lytemps}. Error bars are statistical errors;
systematic errors are small (see \S\ref{syserr}).
}
\end{figure}

 In the top part of Table~\ref{lytemps} we list temperatures
derived from the H-like to He-like line ratios of the same
species (after correction for $N_{\rm H}$ as noted in
Table~\ref{lytemps}) using
theoretical predictions of the same ratios as a function of
temperature extracted from an atomic database computed by
the Astrophysical Plasma Emission Code
\citep[APEC, v1.3:][]{smith01,smith01b}. The temperatures
are computed under the assumption of collisional equilibrium
(see end of this section and \S\ref{methods} for more details)
and are average
values assuming that the plasma is isothermal over the
temperature range over which the respective H-like and
He-like lines are formed. Lines originating from high-Z
elements probe hotter plasma. Since the lines
involved in each ratio are from the same element, these
ratios yield temperatures independently of the elemental
abundances \citep[see, e.g.,][]{abun,ness_vel}. A graphical
representation is given in Fig.~\ref{lrats}, and it can
be seen that the Si ratios probe hotter plasma than the
N ratios. Significantly different temperatures are derived,
indicating that we are dealing with a wide range of
temperatures in the early observations (before
day $\sim 70$). The measurements for the later observations
deliver consistent temperatures (at least within the large
uncertainties), indicating that the plasma could be
characterized by a single temperature by that time. The
temperatures derived from the Mg lines indicate that the
hotter plasma cools until
day $\sim 70$ and remains constant after that time.
A similar behavior can be concluded from the other values
but it is not as clear. For example the temperatures derived
from the Ne lines yield a slight increase, however, the
temperature on day 26.1 may also be anomalously low,
owing to line blends of the Ne\,{\sc ix} lines
\citep{nebr}.
We note that the N lines observed on day 26.1 show a
peculiar behavior in the line shifts and -widths (see
Table~\ref{vlines}), and the derived temperature may
thus not represent the same plasma as the values for the
other times of evolution. For days 39.7-66.9
the cool component cannot be probed because the O and N
lines are outshone by the SSS emission from the WD. The
temperature derived from the O lines for day 39.7 is based
on the fluxes measured by \cite{ness_rsoph} on top of the
SSS continuum. These fluxes may be contaminated by
photoexcitations, but the derived temperature agrees
well with the temperatures derived for the other days.

 In the bottom part of Table~\ref{lytemps} we list
densities derived from the He-like
forbidden-to-intercombination (f/i) line ratios for the
ions O\,{\sc vii} and Ne\,{\sc ix}. These values have
been derived assuming collisional equilibrium according to
the parameterization derived by \cite{gj69}, neglecting
UV radiation; see, e.g., \cite{denspaper} for details. The
method explores density-dependent excitations out of the
upper level of the f line (1s2p\,$^3$S) into that of the
i line (1s2p\,$^3$P). In the low-density limit, all
ions in the 1s2p\,$^3$S state radiate to the
ground (1s$^2$\,$^1$S), giving rise to the f line, while
with increasing density, collisional excitations from the
1s$^2$\,$^3$S state into the 1s2p\,$^3$P state reduce the
f line and increase the i line. However, the
1s2p\,$^3$P-1s2p\,$^3$S transition can
also be induced by UV radiation, whose presence would
mimic a high density if neglected \citep{blum72,ness_cap}.
Especially for the early
spectra we must assume that significant UV contamination
from the WD has to be accounted for. While the
UV intensity can be estimated from IUE observations
of the 1985 outburst \citep{shore96}, the distance
between the X-ray emitting plasma and the UV source
needs to be known in order to quantify the contaminating
effects for the density diagnostics. Since UV radiation
fields, if present, mimic high densities, we treat the
values with great caution, but we can at least conclude
that the density is not higher than any of the values
listed in Table~\ref{lytemps}.

In the next section we present models for which the
assumption of optically thin plasma is made.
As a test of this assumption we measure the line flux
ratio of the Ly$\beta$ (3p-1s) to Ly$\alpha$ (2p-1s) lines of
the H-like ion Ne\,{\sc x} (10.23\,\AA\ and 12.12\,\AA,
respectively), which increases with increasing
optical depth, but also depends on the electron
temperature and the amount of photoelectric absorption
in the line of sight. Theoretical predictions of the same
ratio for an optically thin plasma in collisional equilibrium,
from the atomic databases CHIANTI version 5.2 \citep{landi06}
and APEC, vary between 0.1 and 0.3 within the temperature range
$10^6$ to $10^8$\,K, while for day 13.8 we measure a ratio of
$0.17\,\pm\,0.02$. This is well within the expected range.
We also refer to the tests presented by \cite{nelson07}
who measured the so-called G-ratio of the He-like triplets
of S, Si, and Mg and concluded that the plasma is collisionally
dominated.\\

\begin{table*}[!ht]
\begin{flushleft}
\renewcommand{\arraystretch}{1.1}
\caption{\label{elines}Evolution of emission line fluxes.}
\begin{tabular}{lrccccccccc}
& & flux$^{b}$ & flux$^{b}$& flux$^{b}$ & flux$^{b}$ & flux$^{b}$& flux$^{b}$& flux$^{b}$ \\
Ion & $\lambda_0^{a}$\small{(\AA)}&day 13.81 & day 13.88 & day 26.1 & day 111.7 & day 205.9 & days 204--208$^e$ & day 239.2\\
\hline
Fe\,{\sc xxvi}  & 1.78   & $<144$ &         --            &       --          &    --          &    --         &    --         &    --      \\
Fe\,{\sc xxv}$^c$   & 1.85   & $1037\,\pm\,190$ &     --            &       --          &    --          &    --         &    --         &    --      \\
S\,{\sc xvi}    & 4.73   & $219\,\pm\,22$ & -- &  \multicolumn{2}{c}{day 39.7: $<33.8$}  &    --         &    --         &    --      \\
S\,{\sc xv}$^c$ & 5.04   & $624\,\pm\,84$ & -- &  \multicolumn{2}{c}{day 39.7: $78.1\,\pm\,67.7$} &    --         &    --         &    --      \\
Si\,{\sc xiv}   & 6.18   & $349\,\pm\,13$ & $260\,\pm\,32$  & $138\,\pm\,29$  & $2.7\,\pm\,1.8$  &    --     & $<1.5$ &        --      \\
Si\,{\sc xiii}$^c$  & 6.65   & $620\,\pm\,32$ & $795\,\pm\,112$ & $560\,\pm\,124$ & $10.3\,\pm\,5.6$ &    --     & $<5.1$ &        --      \\
Mg\,{\sc xii}   & 8.42   & $412\,\pm\,11$ & $585\,\pm\,19$  & $269\,\pm\,19$  & $5.1\,\pm\,2.0$  & $<0.3$    & $1.3\,\pm\,0.7$ &  $<0.3$\\
Mg\,{\sc xi}$^c$   & 9.20   & $383\,\pm\,30$ & $369\,\pm\,35$  & $374\,\pm\,47$  & $18.8\,\pm\,7.6$ & $4.5\,\pm\,2.9$ & $4.1\,\pm\,2.4$ &  $2.2\,\pm\,1.1$ \\
Ne\,{\sc x}     & 12.13  & $273\,\pm\,20$ & $309\,\pm\,9.2$   & $309\,\pm\,12$  & $21.6\,\pm\,3.0$ & $3.0\,\pm\,0.7$ & $3.2\,\pm\,0.7$ &  $2.0\,\pm\,0.5$ \\
+Fe\,{\sc xvii} & 12.12&\multicolumn{5}{l}{1.08 times the flux at 12.26\,\AA\ if $\log T<6.9$ otherwise negligible}&\\
Fe\,{\sc xvii}  & 12.26  & $52.8\,\pm\,6.8$   & $52.5\,\pm\,6.6$    & $82.8\,\pm\,9.3$    & $3.2\,\pm\,1.8$  & $<1.6$ & $1.5\,\pm\,0.6$ &   $<0.75$\\
+Fe\,{\sc xxi}  & 12.28&\multicolumn{5}{l}{at $\log T>6.9$ only Fe\,{\sc xxi} otherwise only Fe\,{\sc xvii}}&\\
Ne\,{\sc ix}    & 13.44  & $71.8\,\pm\,8.9$   & $99.8\,\pm\,5.3$    & $148\,\pm\,9.4$   & $19.6\,\pm\,2.8$ & $3.7\,\pm\,0.8$ & $3.4\,\pm\,0.7$ &  $3.1\,\pm\,0.5$ \\
Ne\,{\sc ix} (i)$^{d}$& 13.55 & $19.8\,\pm\,6.8$  & $38.2\,\pm\,4.5$    & $60.3\,\pm\,8.4$    & $3.4\,\pm\,1.8$  & $1.2\,\pm\,0.6$ & $1.0\,\pm\,0.6$ &  $0.7\,\pm\,0.4$ \\
Ne\,{\sc ix} (f)$^{d}$& 13.69 & $33.9\,\pm\,6.8$  & $53.3\,\pm\,4.1$    & $82.3\,\pm\,7.4$    & $10.1\,\pm\,2.1$ & $3.3\,\pm\,0.7$ & $2.8\,\pm\,0.6$ &  $2.4\,\pm\,0.5$ \\
Fe\,{\sc xvii}  & 14.21  & $33.6\,\pm\,7.5$   & $45.5\,\pm\,3.4$    & $59.5\,\pm\,5.6$    & $1.8\,\pm\,1.4$  & $1.6\,\pm\,0.6$ & $1.0\,\pm\,0.5$ &--\\
Fe\,{\sc xvii}  & 15.01  & $67.3\,\pm\,8.1$   & $69.5\,\pm\,3.6$    & $156\,\pm\,7.7$   & $14.3\,\pm\,2.1$ & $3.4\,\pm\,0.7$ & $2.7\,\pm\,0.5$ & $2.7\,\pm\,0.5$ \\
O\,{\sc viii}   & 18.97  & $84.2\,\pm\,19$  & $92.5\,\pm\,3.1$    & $186\,\pm\,13$  & $33.5\,\pm\,2.8$ & $7.7\,\pm\,1.4$ & $5.3\,\pm\,0.6$ &  $5.2\,\pm\,0.5$ \\
O\,{\sc vii}    & 21.60  &     --               & $14.8\,\pm\,1.9$     & $29.5\,\pm\,3.1$    & $6.8\,\pm\,1.8$  & $2.4\,\pm\,0.5$ & $2.1\,\pm\,0.6$ &  $0.9\,\pm\,0.3$ \\
O\,{\sc vii} (i)$^{d}$& 21.80 &     --              & $<1.5$     & $13.6\,\pm\,2.4$    & $1.3\,\pm\,1.3$  & $1.4\,\pm\,0.4$ & $<0.5$ &  $0.6\,\pm\,0.3$ \\
O\,{\sc vii} (f)$^{d}$& 22.10 &     --              & $6.4\,\pm\,1.3$     & $33.4\,\pm\,3.2$    & $5.4\,\pm\,1.7$  & $2.0\,\pm\,0.5$ & $0.8\,\pm\,0.6$ &  $1.0\,\pm\,0.3$ \\
N\,{\sc vii}    & 24.78  &     --               & $43.0\,\pm\,1.8$    & $259\,\pm\,12.8$  & $15.7\,\pm\,2.3$ & $6.6\,\pm\,0.7$ & $3.9\,\pm\,0.6$ &  $4.0\,\pm\,0.5$ \\
N\,{\sc vi}     & 28.78  &     --           &    $2.3\,\pm\,0.6$ &    $35.4\,\pm\,3.5$ & $1.8\,\pm\,1.1$  &  $<0.4$ &   $<1.5$ &  $0.4\,\pm\,0.2$\\
C\,{\sc vi}     & 33.74 & -- & $<0.6$ & $<0.5$ & $<1.4$ & $<0.3$ & $<0.9$ & $<0.1$\\

\hline
\multicolumn{9}{l}{Temperature-sensitive line ratios}\\
Fe\,{\sc xxvi}/Fe\,{\sc xxv} && $<0.2$&--&--&--&--&--&--\\
S\,{\sc xvi}/S\,{\sc xv} &&   $0.35\,\pm\,0.06$&--& \multicolumn{2}{c}{day 39.7: $<3.25$}&--&--&--\\
Si\,{\sc xiv}/Si\,{\sc xiii}& &$0.56\,\pm\,0.04$ & $0.33\,\pm\,0.06$ & $0.25\,\pm\,0.08$ & $0.26\,\pm\,0.23$ & -- & -- &--\\
Mg\,{\sc xii}/Mg\,{\sc xi}& & $1.08\,\pm\,0.09$ & $1.59\,\pm\,0.16$ & $0.72\,\pm\,0.10$ & $0.27\,\pm\,0.15$ & $<0.2$ & $0.32\,\pm\,0.25$ & $<0.3$ \\
Ne\,{\sc ix}/Ne\,{\sc x} &&$3.80\,\pm\,0.54$ & $3.09\,\pm\,0.19$ & $2.10\,\pm\,0.16$ & $1.10\,\pm\,0.22$ & $0.81\,\pm\,0.26$ & $0.94\,\pm\,0.28$ & $0.65\,\pm\,0.19$ \\
O\,{\sc viii}/O\,{\sc vii} &&-- & $6.25\,\pm\,0.83$ & $6.31\,\pm\,0.79$ & $4.93\,\pm\,1.37$ & $3.21\,\pm\,0.89$ & $2.52\,\pm\,0.78$ & $5.78\,\pm\,2.00$ \\
N\,{\sc vii}/N\,{\sc vi} &&-- & $18.70\,\pm\,4.94$ & $7.32\,\pm\,0.81$ & $8.72\,\pm\,5.48$ & $>14.8$ & $>2.2$ & $10.00\,\pm\,5.15$ \\
\hline
\multicolumn{9}{l}{Density-sensitive line ratios}\\
Ne\,{\sc ix} (f/i) && $1.7\,\pm\,0.7$ & $1.4\,\pm\,0.2$ & $1.4\,\pm\,0.2$ & $3.0\,\pm\,1.7$ & $2.8\,\pm\,1.5$ & $2.8\,\pm\,1.8$ & $3.4\,\pm\,2.1$\\
O\,{\sc vii} (f/i) && -- & $>4.3$ & $2.5\,\pm\,0.5$ & $>4.2$ & $1.4\,\pm\,0.5$ & $>4.0$ & $1.7\,\pm\,1.0$\\
\hline
\end{tabular}
Uncertainties and upper limits are 68.3\%. Additional systematic
uncertainties are discussed in \S\ref{syserr}.\ $\bullet$\
$^{a}$rest wavelengths $\bullet\ ^{b}10^{-14}$\,erg\,cm$^{-2}$\,s$^{-1}$
$\bullet\ ^{c}$sum of three lines
$\bullet\ ^{d}$intersystem line
$\bullet\ ^e$Sum of Chandra spectra taken between days 203.3 and 208.2 (see
Table~\ref{tab1}).
\renewcommand{\arraystretch}{1}
\end{flushleft}
\end{table*}

\begin{table*}[!ht]
\begin{flushleft}
\renewcommand{\arraystretch}{1.1}
\caption{\label{lytemps}Temperatures and Densities derived from H-like to He-like line ratios}
\begin{tabular}{lrccccccccc}
Ion & day 13.81 & day 26.1 & day 39.7 & day 54 & day 66.9 & day 111.7 & days 204--208$^a$ & day 239.2\\
\hline
\multicolumn{9}{l}{$\log T$ in K, after correction of line ratios
for $N_{\rm H}$ with values $5\times 10^{21}$\,cm$^{-2}$ for
days 13.8 and 26.1 and $2.4\times 10^{21}$\,cm$^{-2}$ for the
rest.}\\
Fe & $<7.8$&--&--&--&--&--&--\\
S & $7.21^{+0.02}_{-0.20}$ & -- & $<7.61$&--&--&--&--\\
Si & $7.09\,\pm\,0.01$ & $6.98\,\pm\,0.03$ & $7.02\,\pm\,0.02$ & $7.01^{+0.05}_{-0.1}$ & $7.05^{+0.08}_{-0.14}$ & $7.00^{+0.08}_{-0.19}$ & -- & -- \\
Mg &$6.96\,\pm\,0.01$ & $6.91\,\pm\,0.02$ & $6.87\,\pm\,0.02$ & $6.85\,\pm\,0.02$ & $6.81\,\pm\,0.03$ & $6.80^{+0.04}_{-0.1}$ & $6.81^{+0.07}_{-0.16}$ & $<6.8$ \\
Ne &$6.81\,\pm\,0.02$ & $6.72\,\pm\,0.01$ &$6.74\,\pm\,0.02$ & $6.77\,\pm\,0.03$ & $6.79^{+0.05}_{-0.1}$& $6.65\,\pm\,0.03$ & $6.63\,\pm\,0.04$ & $6.59\,\pm\,0.03$ \\
O & $6.46\,\pm\,0.02$ & $6.46\,\pm\,0.02$ & $6.42\,\pm\,0.03$ &--&--& $6.52\,\pm\,0.04$ & $6.43\,\pm\,0.04$ & $6.55^{+0.05}_{-0.1}$ \\
N & $6.36\,\pm\,0.04$ & $6.23\,\pm\,0.01$ &--&--&--& $6.40^{+0.08}_{-0.15}$ & $>6.5$ & $6.42\,\pm\,0.11$ \\
\hline
\multicolumn{9}{l}{Densities, $\log n_e$ in cm$^{-3}$, from He-like triplet ratios assuming no UV illumination (see text \S\ref{lfluxmeas})}\\
O\,{\sc vii} & $<8$ & $10.3^{+0.7}_{-0.2}$ & -- & -- & -- & $<8$ & $10.7^{+0.2}_{-0.8}$ & $10.6^{+0.2}_{-0.8}$\\
Ne\,{\sc ix} & $11.7\,\pm\,0.4$ & $11.9^{+0.2}_{-0.4}$ & -- & -- & -- & $>10.3$ & $>10.8$ & $>9$\\
\hline
\end{tabular}
$^a$Sum of Chandra spectra taken between days 203.3 and 208.2.
The results are consistent with {\it XMM}
observations taken day 205.9.
\renewcommand{\arraystretch}{1}
\end{flushleft}
\end{table*}

\section{Analysis}
\label{anal}

 For the interpretation of the observations described above
we use two model approaches. First we compute multi-temperature
plasma models with the fitting package {\sc xspec} \citep{xspec}.
We use the atomic data computed by APEC which are similar to
the MEKAL database, and the results can be compared to those
given by \cite{bode06} for earlier X-ray observations.
Next, we use the measured emission line fluxes from
Table~\ref{elines} to construct a model of a smooth temperature
distribution. This model allows us to determine relative
abundances.

\subsection{Description of methods and model assumptions}
\label{methods}

 For our modeling of the X-ray spectra of the shock we assume:
(1) that all emission originates
from the same volume with the same abundances, (2) that the
plasma is in a collisional equilibrium, and (3) that it is
optically thin.

The first assumption is implicit in all spectral analyses
in X-rays unless spatial resolution is available. Although
there is likely stratification to some extent in the
emitting environment, we have no basis on which we can
develop more refined models. We further have to assume a
uniform plasma, which implies that interstellar and
circumstellar absorption can only be modelled with
a single absorption component.

The line ratios used in the end of \S\ref{lfluxmeas} support the other
two assumptions. In a collisional plasma, all temperatures
are kinetic temperatures, derived from the distribution of
velocities, which is commonly assumed to be Maxwellian. While
in a shocked plasma collisions are the main energy source
for all atomic transitions, the assumption of an equilibrium
is not necessarily valid. Note that
rapid recombination can lead to non-equilibrium conditions,
since recombination into excited states leads to
an overpopulation of upper levels and consequently to
excessively high fluxes in certain emission lines. Nevertheless,
we base our analysis on the assumption of equilibrium
conditions and discuss the implications of this assumption
where relevant.

The third assumption implies that all emission that is produced
in the collisional plasma escapes unaltered. However, lines
with high oscillator strengths may be reabsorbed within the
plasma and reemitted in a different direction (resonant
line scattering). Depending on the plasma geometry, resonance
lines may be stronger or weaker compared to optically thin
plasma, since resonance line photons can be scattered out of
the line of sight or into the line of sight. In a spherical
geometry the processes cancel out and no effects from
resonant line scattering are detectable. Ways to detect
resonant line scattering are discussed by \cite{ness_opt}.
The emission line fluxes presented in \S\ref{lfluxmeas} show
no signatures of resonant line scattering.

In a collisional plasma the brightness of a source is
expressed in terms of the volume emission measure,
$VEM$, which is a measure of the intensity per unit volume
(in cm$^{-3}$). The $VEM$
is defined as $VEM=\int n_{\rm e}^2{\rm d}V$ with
$n_{\rm e}$ the electron density and $V$ the emitting
volume. A given value of $VEM$ is thus proportional to the
emitting volume; however, volumes can only be determined
from independent density measurements which are difficult
to obtain from X-ray spectra (see \S\ref{lfluxmeas}).

The volume emission measure as a function of temperature,
$T$, is called the emission measure distribution (EMD). A given
EMD is a model that allows the calculation of continuum
emission by bremsstrahlung and emission line fluxes, which
together form a predicted X-ray spectrum that can be
compared to an observed spectrum. In order to predict the
line flux for a line at wavelength $\lambda$, one needs the
respective line contribution function with temperature $T$
\begin{equation}
\label{gte}
G_\lambda(T)=0.83\frac{h\,c}{\lambda}\frac{n_u(T)}{n_I(T)n_{\rm e}(T)}\frac{n_I(T)}{n_E}\frac{n_E}{n_{\rm H}}A_{ul}\frac{1}{4\pi d^2}
\end{equation}
with $h$ Planck's constant, $c$ the speed of light and $\lambda$
the wavelength of the line. The number densities are given with
$n$ with subscripts $u$ for ions in the upper level in the
transition, $I$ for the ionization stage in which the transition
occurs, $E$ for the element giving rise to the transition,
and $n_e$ is the electron density.
The ratio $n_I(T)/n_E$ is the ionization balance, and calculations
of $n_I(T)/n_E$ as a function of temperature have been presented
by \cite{mazzotta}. The ratio $n_E/n_{\rm H}$ represents
the absolute elemental abundance, and $A_{ul}$ is the Einstein
A-value. With a given EMD (which can also be written as
$VEM(T)$) the predicted line fluxes are
\begin{equation}
\label{fpred}
 F_\lambda=\int VEM(T)\times G_\lambda(T) \times {\rm d}T\,.
\end{equation}
An EMD can be defined as one or more isothermal components
(\S\ref{xspecsect}) or as a continuous function of temperature
(\S\ref{lfluxes}).

When fitting models using {\sc xspec} (\S\ref{xspecsect}) all
information that is available in the atomic database is used
to constrain the models. We make use of a second approach
(\S\ref{lfluxes}) selecting only the most reliable
atomic data, and constrain the models by fitting the
measured line fluxes rather than the entire spectrum.
Each approach has strengths and weaknesses, and we compare
the two approaches in \S\ref{cmpmodels}.

 In \S\ref{syserr} we discuss estimates of systematic
uncertainties in addition to the given statistical
uncertainties.

\subsection{Multi-temperature plasma models}
\label{xspecsect}

\begin{table*}[!ht]
\begin{flushleft}
\renewcommand{\arraystretch}{1.1}
\caption{\label{xspec}Model parameters from multi-$T$ APEC
models}
\begin{tabular}{lccccccc}
\hline
        & day 13.8 & day 26.1$^b$ & day 26.1$^c$ & day 39.7 & day 66.9 & day 111.7 & day 239.2\\
k$T_1^a$ & $4.19\,\pm\,0.11$ & $1.80\,\pm\,0.24$ & $1.91\,\pm\,0.07$ & $1.88\,\pm\,0.13$   & $74.4^{+5.5}_{-62}$ & $0.74\,\pm\,0.06$ & $<79.9$\\
$\log(T)^a$ & $7.69\,\pm\,0.01$ & $7.32\,\pm\,0.06$ & $7.35\,\pm\,0.02$ & $7.34\,\pm\,0.03$ & $8.93^{+0.01}_{-8.13}$ & $6.93\,\pm\,0.04$ & $<8.97$\\
$\log(EM_1)^a$  & $58.02\,\pm\,0.01$ & $57.48\,\pm\,0.04$ & $57.51\,\pm\,0.02$ & $57.23\,\pm\,0.03$ & $56.6\,\pm\,0.14$ & $56.15\,\pm\,0.09$ & $<54.04$\\
k$T_2^a$ & $0.74\,\pm\,0.01$ & $0.74\,\pm\,0.02$ & $0.74\,\pm\,0.01$ & $0.60\,\pm\,0.01$ & $0.66^{+0.13}_{-0.09}$ & $0.35\,\pm\,0.03$ & $0.37\,\pm\,0.02$\\
 $\log(T)^a$ & $6.93\,\pm\,0.01$ & $6.93\,\pm\,0.02$ & $6.93\,\pm\,0.01$ & $6.84\,\pm\,0.01$  & $6.88^{+0.08}_{-0.07}$ & $6.61\,\pm\,0.11$ & $6.63\,\pm\,0.03$\\
$\log(EM_2)^a$& $57.76\,\pm\,0.01$  & $57.69\,\pm\,0.05$  & $57.67\,\pm\,0.04$ & $57.44\,\pm\,0.02$  & $56.81\,\pm\,0.20$ & $55.21\,\pm\,0.09$ & $55.65\,\pm\,0.03$\\ 
k$T_3^a$ & $0.30\,\pm\,0.01$ & $0.37\,\pm\,0.03$ & $0.38\,\pm\,0.02$ & $0.11\,\pm\,0.01$ & $0.14\,\pm\,0.06$ & $0.12\,\pm\,0.03$ & $0.11\,\pm\,0.02$\\
 $\log(T)^a$ & $6.55\,\pm\,0.01$ & $6.64\,\pm\,0.96$ & $6.64\,\pm\,0.03$ & $6.10\,\pm\,0.06$ & $6.21\,\pm\,0.24$& $6.14\,\pm\,0.11$&$6.12\,\pm\,0.10$\\
$\log(EM_3)^a$& $57.46\,\pm\,0.02$ & $57.60\,\pm\,0.09$ & $57.61\,\pm\,0.05$ & $58.96^{+0.7}_{-0.3}$ & $58.88\,\pm\,1.21$ & $55.88\,\pm\,0.25$ & $55.32\,\pm\,0.45$\\ 
$N_{\rm H}$ & $6.95\,\pm\,0.35$  & $5.59\,\pm\,0.02$ & $5.56\,\pm\,0.13$ & 2.4 & 2.4 & 2.4 & 2.4\\
$\chi^2_{\rm red},\,dof^{\,d}$& 0.67,\,{\it 21317} & 1.70,\,{\it 2939} & 1.73,\,{\it 3151} & (0.65,\,{\it 1928})$^e$ & (0.18,\,{\it 1294})$^e$ & (0.21,\,{\it 10088})$^e$ & (0.59,\,{\it 5228})$^e$\\
\hline
\end{tabular}

$^a$Units: k$T$ in keV, $T$ in K, $EM$ in cm$^{-3}$, $N_{\rm H}$
in $10^{21}$\,cm$^{-2}$
$\bullet\ ^b$Only RGS data\ $\bullet\ ^c$RGS and MOS1 data
simultaneously\ $\bullet\ ^d$degrees of freedom\ 
$\bullet\ ^e$after iteration with C-statistics\ 
$\bullet\ ^f$fixed at values
from day 13.8
\renewcommand{\arraystretch}{1}
\end{flushleft}
\end{table*}

 We construct spectral models using the X-ray fitting
package {\sc xspec} version 11.3.2ag \citep{xspec} which
combines all information in a given atomic database and
generates count spectra to be compared to the observed
count spectra. We chose multi-temperature APEC models
comprising the sum of $n$ independent isothermal
components with variable abundances relative to the
solar values listed by \cite{agrev89}. Line fluxes are
obtained applying Eq.~\ref{fpred} using the ionization
balance by \cite{mazzotta}, and the plasma density is
assumed constant at $\log n_e=0$ \citep{smith01}.
We vary only
the abundances of elements which produce strong emission
lines in the respective spectra (see Figs.~\ref{day13.8} to
\ref{day111.7}) and assume solar abundances
for the other elements. We correct for interstellar
absorption using the {\tt tbabs} module developed
by \cite{wilms00} and allow the neutral hydrogen column
density, $N_{\rm H}$, to vary
only for the first two observations. For the later
observations we fix $N_{\rm H}$ at the interstellar value
of $N_{\rm H}=2.4\times10^{21}$\,cm$^{-2}$.
The effects of $N_{\rm H}$ have a stronger influence on the
cooler component and may affect the abundances of elements
whose lines are formed at longer wavelength (i.e., nitrogen
and oxygen).

\begin{figure*}[!ht]
\resizebox{\hsize}{!}{\includegraphics{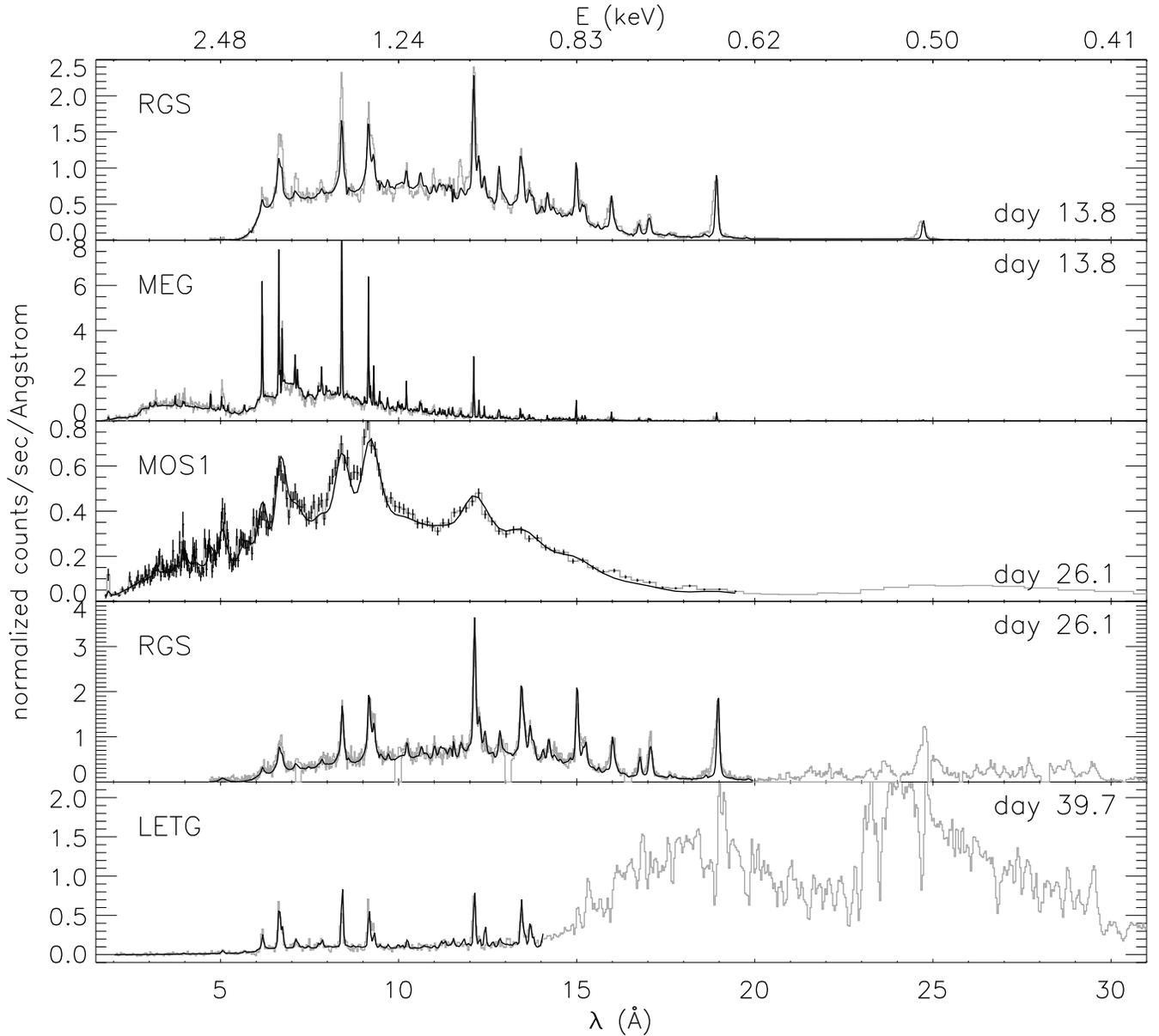}}
\caption{\label{mepic}Best-fit APEC models (from top
to bottom) to the
\xmm/RGS and \chandra/MEG spectra on day 13.8
\xmm/MOS1 and RGS spectra taken on day 26.1, and the
\chandra/LETG spectrum taken on day 39.7. For day 26.1
the soft component longwards of 20\,\AA\ was excluded from
the fit, and for day 39.7 the SSS spectrum had to be excluded.
}
\end{figure*}

 We use the optimization procedures provided by the
{\sc xspec} programme to obtain best fits and the {\sc xspec}
command {\tt error} to calculate the parameter uncertainties,
yielding 90-per cent uncertainties. This command steps through
a range for a given parameter, and in the course of this
process further improvements of the fit can be found.
The uncertainties returned by the {\tt error} command are
only statistical uncertainties that describe the precision
of the measurement but not necessarily the accuracy
of the respective parameters (see \S\ref{syserr}). The
results are listed in Table~\ref{xspec}, and some of the
corresponding models are shown in Fig.~\ref{mepic}. The
elemental abundances are only varied for day 13.8, because
this is the best dataset. The values relative to
\cite{agrev89} adopted for all observations are given in
Table~\ref{abutab}, middle column.

 While the true EMD is most likely a continuous distribution,
we use multi-temperature models. The only continuous EMD
models to chose from in {\sc xspec} are Chebyshev polynomials
of no more than six
orders and are not constrained to be positive. We regard
these EMDs as not sufficient for our purposes and therefore
use the more standard multi-$T$ models. The number of free
parameters, and thus the number of temperature components,
has to be chosen to be as small as possible while still
achieving a good fit. The spectra taken on day 13.8 have high
statistical quality, and a 3-temperature (=3-$T$) model yields
significantly better reproduction of the data than 2-temperature
models. With variable abundances, no fourth temperature component
is required to improve the fit. The APEC model has
a redshift parameter that we allow to vary in order to
account for line shifts (see \S\ref{lprofiles}),
however, we cannot account for line broadening in excess
of the instrumental line broadening function. While the
model could be folded with a Gaussian with variable width,
this is computationally expensive and unfeasible with the
given high number of spectral bins and free parameters to be
iterated. The long-wavelength lines and the associated
elemental abundances (particularly of N and O) may thus
be poorly determined, and our second approach (\S\ref{lfluxes})
is more reliable for the abundances of these elements.
Meanwhile, the lines at
shorter wavelengths are not broadened by as much, and the
abundances of the other elements are less affected.
 We fit the model
simultaneously to the HEG and MEG spectra plus the RGS1
and RGS2 spectra (top panels in Fig.~\ref{mepic}).
Our model agrees better with the RGS data (top panel) than
that presented by \cite{nelson07}, fig.~3, who have not
varied the abundances but used four temperature components.
No formal value of $\chi^2$ was given, but visual inspection
clearly shows that their model does not reproduce the
RGS spectrum. This demonstrates that the effects from
non-solar abundances are indeed detectable.

 Since on day 26.1 the source was still bright, the observed
spectra are also of high statistical quality, and 3-$T$
models are better than 2-temperature models. Since we expect no
detectable changes in the composition, we use the elemental
abundances found from the day 13.8 observations for this and
the following datasets. We discard all
spectral bins longward of 20\,\AA\ because the emission does not
originate from the shock (third and fourth panels in
Fig.~\ref{mepic}). The N abundance is now less certain because
the strong N\,{\sc vii} line at 24.78\,\AA\ is excluded.
We concentrate on the RGS spectra but
also compute a model including the MOS1 spectra which
are sensitive at higher energies and are thus suited
to constrain the hotter component. As can be seen from
Table~\ref{xspec}, the model parameters are identical, and
only the
uncertainties of the model with the MOS1 data included
are smaller. The hot component is thus detectable with
the RGS alone. 

Since the spectra taken on days 39.7-66.9 are compromised
by the SSS emission (bottom panel of Fig.~\ref{mepic})
the X-ray emission from the low-temperature shocked plasma
cannot be probed. The spectrum shortwards of $\sim 14$\,\AA\
is not well enough exposed to require three temperature
components. However, in order to compare the results, we allow
three temperature components with the option that the fitting
procedure can assign small values of emission measure to
those components that are not detectable. We caution, however,
that the hottest temperature component that probes the
bremsstrahlung continuum may be overestimated due to
systematic uncertainties in the instrumental background
(see \S\ref{syserr}).
If the theoretical continuum is of the same order as the
noise in the background, arbitrarily high temperatures
may result which will have to be treated with caution.
Since the spectra taken after day 26.1 contain many bins with
low counts, we use C-statistics \citep{cash79}. This approach
is the based on the maximum likelihood (ML) method
described in the appendix. We calculate
a formal value of $\chi^2_{\rm red}$ after fitting with
{\tt cstat} for comparison with the other fits. We use the
errors on the count rates from the extracted spectra.\\

\begin{figure}[!ht]
\resizebox{\hsize}{!}{\includegraphics{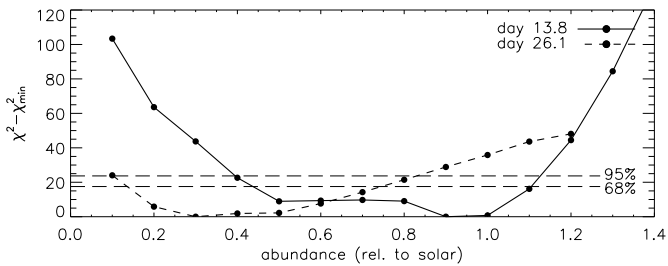}}
\caption{\label{absabu}Elemental abundances relative to solar
\citep{agrev89} from APEC fits to the spectra of days
13.8 and 26.1. Each grid point is the result of
iterating all free parameters with the absolute abundance
fixed at the respective grid value. Plotted are the changes in
$\chi^2$ compared to the best fit. The dashed horizontal
lines mark the 68-\% and 95-\% confidence ranges.
}
\end{figure}

 Because the spectra on days 13.8 and 26.2 are of such high
quality, we investigate absolute abundances using these
two datasets
(Fig.~\ref{absabu}). Although no hydrogen lines are present in 
the X-ray range, the absolute abundances can be determined
from the strength of the continuum relative to the lines.
The brightness of the continuum depends on the number of free
electrons which, in an ionized plasma, scales with the
hydrogen abundance. We thus need spectra with sufficient
continuum emission. We step
through a grid of (fixed) abundances and fit the remaining
parameters to minimize $\chi^2$. In Fig.~\ref{absabu}
we show the relative changes in $\chi^2$ for each grid
point in comparison with the 68-\% and 95-\% confidence ranges
(for 14 free parameters). While for day 13.8, solar abundances
are preferred, the spectrum taken on day 26.1 suggests a
somewhat lower metallicity, but from the confidence intervals one
can see that their determination is highly uncertain.
We therefore fix the absolute abundance at
solar values and concentrate on the relative abundances.

The final model parameters are summarized in Table~\ref{xspec}.
From top to bottom we list $\log T$ and $\log VEM$ for each
component, value of $N_{\rm H}$, and $\chi^2_{\rm red}$ with
number of degrees of freedom ($dof$). The elemental abundances
relative to solar \citep{agrev89} as determined from the day 13.8
dataset are given in Table~\ref{abutab} and have been used for
all models.

\begin{figure}[!ht]
\resizebox{\hsize}{!}{\includegraphics{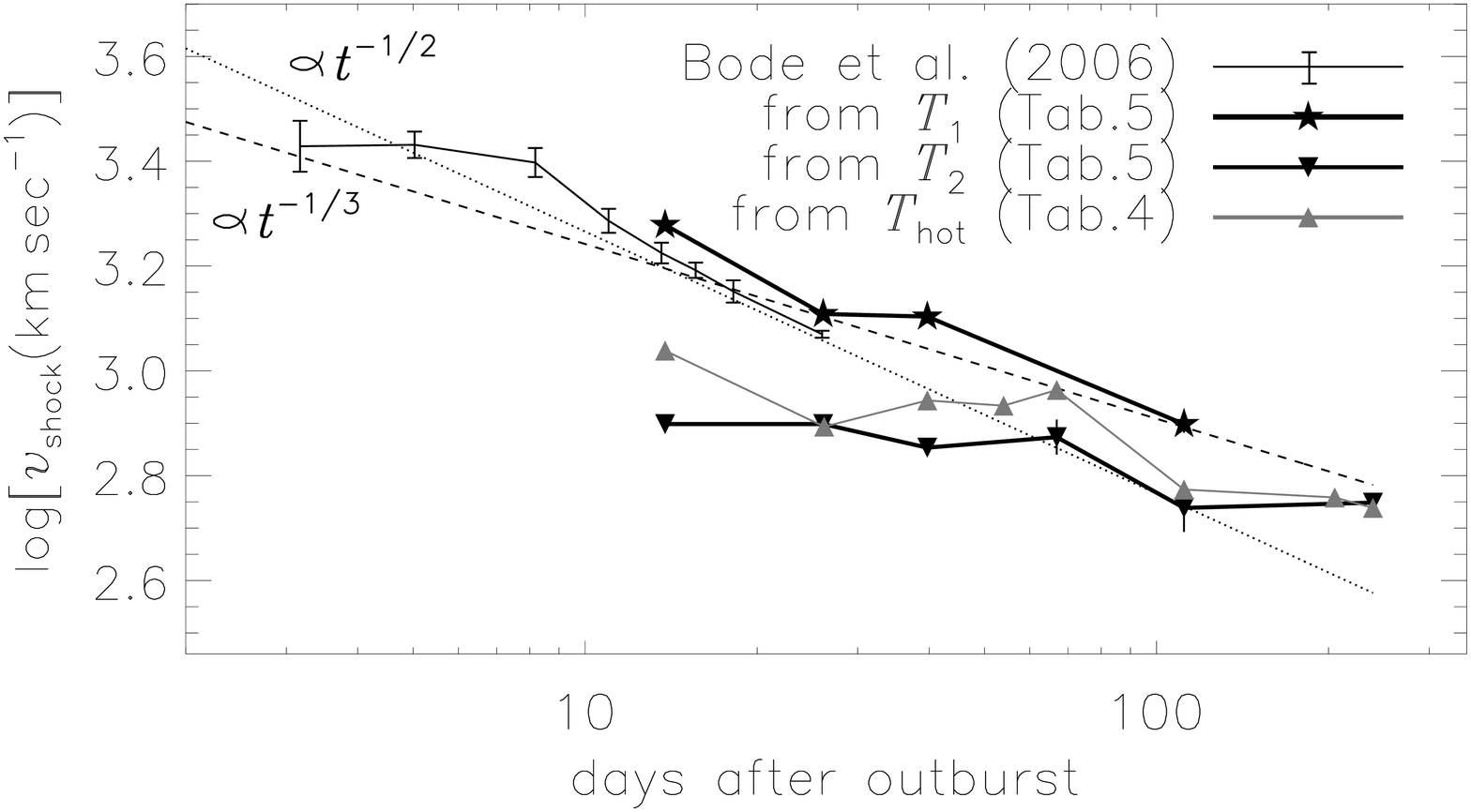}}
\caption{\label{evol_v}Evolution of the shock velocity obtained
from the temperatures found by \cite{bode06}, from the
temperatures of the first and second components of the APEC
models (Table~\ref{xspec}), and the highest temperatures measured
from line ratios (Table~\ref{lytemps}).
The dashed and dotted lines indicate the expected power law decay
in an adiabatic plasma ($t^{-1/3}$) and
in a radiatively cooling plasma ($t^{-1/2}$), respectively.
The error bars are statistical errors. For additional
systematic errors see \S\ref{syserr}.
}
\end{figure}

In the same way as \cite{bode06} we compute shock velocities,
$v_{\rm shock}$, from the temperatures of the first and second
component of the APEC models (see Table~\ref{xspec}), and from
the highest temperatures found from line ratios (Table~\ref{lytemps}).
In Fig.~\ref{evol_v} we compare these results with those from
\cite{bode06}. The dotted and dashed lines indicate the expected
evolution for a radiatively cooling plasma and an adiabatic plasma,
respectively \citep{bodekahn85}. The velocities derived from
$T_1$ of our 3-$T$ APEC models for days 13.8 and 26.1 are slightly 
higher than the Bode values. The reason is that the 1-$T$ models
used by \cite{bode06} are an average of all temperature
components, accounting for some of the cooler plasma that in
our 3-$T$ models are accounted for by the two cooler components.
The evolution of $T_1$ follows the same trend as observed by
\cite{bode06}. After day 26.2, the hottest component is much
fainter, and $T_1$ is less certain. In Figs.~\ref{day39.7} and
\ref{day111.7} one can see that the continuum emission level
is significantly lower than that seen in Figs.~\ref{day13.8}
and \ref{day26.1}. Since the parameters of the hottest
temperature component are dominated by the continuum,
systematic uncertainties from background noise have a
stronger effect, and the velocities derived from the hottest
temperature components may be overestimated for the observations
taken after day 26.1 (see \S\ref{syserr}).

The second plasma component is very similar to the values
derived from the line ratios (Table~\ref{lytemps}), but these
curves follow a different trend than the hottest component.
For the observations of days
13.8 and 26.1, the line ratios yield much lower velocities
than those from the APEC models. We attribute this
difference to the stronger continuum observed for these
two days. Since the continuum is dominated by the hottest
plasma, the hottest temperature in the APEC models is driven
by the continuum which, during the early observations,
reflects a higher temperature than any of the emission
lines can probe. Meanwhile, the emission lines can probe
the structure of the temperature distribution better, and
in the next section we describe an approach that focuses on
a few selected emission lines.

\subsection{Emission measure modeling}
\label{lfluxes}

\begin{table}[!ht]
\begin{flushleft}
\renewcommand{\arraystretch}{1.1}
\caption{\label{fluxpred}Comparison of line flux measurements with predictions for day 13.8}
\begin{tabular}{lccc}
\hline
Ion$^a$ & $\lambda$ & $F_{\rm meas}^b$ & $\frac{F_{\rm meas}}{F_{\rm pred}}^c$ \\
\hline
{\bf N\,{\boldmath \scriptsize{VI}}} & 28.79 & 543 & $0.93\,\pm\,0.23$\\
{\bf N\,{\boldmath \scriptsize{VII}}} & 24.78 & 1514 & $1.03\,\pm\,0.04$\\
{\bf O\,{\boldmath \scriptsize{VII}}} & 21.60 & 1200 & $1.13\,\pm\,0.14$\\
O\,{\scriptsize VII} & 22.10 & 686 & $1.00\,\pm\,0.20$\\
{\bf O\,{\boldmath \scriptsize{VIII}}} & 18.97 & 1871 & $0.86\,\pm\,0.19$\\
{\bf Ne\,{\boldmath \scriptsize{IX}}} & 13.45 & 358 & $0.95\,\pm\,0.12$\\
Ne\,{\scriptsize IX} & 13.70 & 183 & $0.96\,\pm\,0.19$\\
{\bf Ne\,{\boldmath \scriptsize{X}}} & 12.14 & 927 & $0.99\,\pm\,0.07$\\
{\bf Mg\,{\boldmath \scriptsize{XI}}} & 9.17 & 705 & $1.03\,\pm\,0.08$\\
{\bf Mg\,{\boldmath \scriptsize{XII}}} & 8.42 & 669 & $0.99\,\pm\,0.03$\\
{\bf Si\,{\boldmath \scriptsize{XIII}}} & 6.65 & 816 & $1.05\,\pm\,0.05$\\
{\bf Si\,{\boldmath \scriptsize{XIV}}} & 6.19 & 438 & $0.94\,\pm\,0.03$\\
{\bf S\,{\boldmath \scriptsize{XV}}} & 5.04 & 708 & $1.10\,\pm\,0.15$\\
{\bf S\,{\boldmath \scriptsize{XVI}}} & 4.73 & 245 & $0.72\,\pm\,0.07$\\
Fe\,{\scriptsize XVII} & 12.26 & 65.0 & $1.13\,\pm\,0.61$\\
Fe\,{\scriptsize XVII} & 15.02 & 448 & $0.85\,\pm\,0.10$\\
Fe\,{\scriptsize XVIII} & 14.21 & 216 & $1.17\,\pm\,0.26$\\
Fe\,{\scriptsize XX} & 12.83 & 69.8 & $0.69\,\pm\,0.39$\\
Fe\,{\scriptsize XX} & 12.85 & 97.7 & $1.08\,\pm\,0.44$\\
Fe\,{\scriptsize XXI} & 12.28 & 174 & $0.96\,\pm\,0.39$\\
Fe\,{\scriptsize XXII} & 11.77 & 92.5 & $1.07\,\pm\,0.20$\\
Fe\,{\scriptsize XXIII} & 10.98 & 69.6 & $1.33\,\pm\,0.26$\\
Fe\,{\scriptsize XXIII} & 11.74 & 99.8 & $1.49\,\pm\,0.25$\\
Fe\,{\scriptsize XXIV} & 11.17 & 46.8 & $1.46\,\pm\,0.41$\\
Fe\,{\scriptsize XXV} & 1.85 & 1045 & $29.4\,\pm\,5.40$\\
Fe\,{\scriptsize XXVI} & 1.78 & $<145$ & $<5.19$\\
\hline
\end{tabular}

$\bullet\ ^a$Lines used in deriving the mean EMD are given in bold face.\
$\bullet\ ^b$Fluxes from Table~\ref{elines} in $10^{-14}$~erg~cm$^{-2}$~s$^{-1}$, corrected for absorption assuming
$N_{\rm H}=5\times10^{21}$\,cm$^{-2}$\
$\bullet\ ^{c\,}$Predicted from the derived EMD, assuming constant pressure $\log P_{\rm e}=13.0$\
\renewcommand{\arraystretch}{1}
\end{flushleft}
\end{table}

\begin{table}[!ht]
\begin{flushleft}
\renewcommand{\arraystretch}{1.1}
\caption{\label{abutab}Elemental abundances relative to solar}
\begin{tabular}{lrrr}
\hline
&\multicolumn{2}{c}{day 13.8}&day 111.7\\
&EMD model&APEC model&EMD model\\
\hline
\multicolumn{4}{l}{correction factors}\\
O&1.0&$0.77\,\pm\,0.03$&1.0\\
N&$6.94\,\pm\,0.44$&$7.12\,\pm\,0.70$&$7.54^{+12}_{-5}$\\
Ne&$1.39\,\pm\,0.11$&$0.68\,\pm\,0.03$&$1.36\,\pm\,0.16$\\
Mg&$2.29\,\pm\,0.08$&$1.06\,\pm\,0.03$&$1.91\,\pm\,0.70$\\
Si&$2.04\,\pm\,0.08$&$0.88\,\pm\,0.03$&$1.62\,\pm\,1.0$\\
S&$3.16\,\pm\,0.21$&$0.69\,\pm\,0.12$&--\\
Fe&$0.50\,\pm\,0.04$&$0.29\,\pm\,0.01$&$0.43\,\pm\,0.10$\\
\hline
\multicolumn{4}{l}{relative to oxygen}\\
$[$N/O$]$&$0.84\,\pm\,0.03$&$0.95\,\pm\,0.05$&$0.88^{+0.41}_{-0.47}$\\
$[$Ne/O$]$&$0.14\,\pm\,0.04$&$0.04\,\pm\,0.03$&$0.13\,\pm\,0.05$\\
$[$Mg/O$]$&$0.36\,\pm\,0.02$&$0.24\,\pm\,0.02$&$0.28\,\pm\,0.20$\\
$[$Si/O$]$&$0.31\,\pm\,0.02$&$0.16\,\pm\,0.02$&$0.21\,\pm\,0.42$\\
$[$S/O$]$&$0.50\,\pm\,0.03$&$0.17\,\pm\,0.08$&--\\
$[$Fe/O$]$&$-0.30\,\pm\,0.03$&$-0.50\,\pm\,0.03$&$-0.37\,\pm\,0.11$\\
\hline
\multicolumn{4}{l}{relative to iron}\\
$[$N/Fe$]$&$1.14\,\pm\,0.05$&$1.43\,\pm\,0.07$&$1.24\,\pm\,0.58$\\
$[$O/Fe$]$&$0.30\,\pm\,0.05$&$0.50\,\pm\,0.03$&$0.37\,\pm\,0.09$\\
$[$Ne/Fe$]$&$0.45\,\pm\,0.05$&$0.53\,\pm\,0.05$&$0.47\,\pm\,0.15$\\
$[$Mg/Fe$]$&$0.66\,\pm\,0.04$&$0.73\,\pm\,0.03$&$065\,\pm\,0.23$\\
$[$Si/Fe$]$&$0.61\,\pm\,0.04$&$0.65\,\pm\,0.04$&$0.58\,\pm\,0.35$\\
$[$S/Fe$]$&$0.80\,\pm\,0.05$&$0.67\,\pm\,0.12$&--\\
\hline
\end{tabular}

all values relative to \cite{grev}
\renewcommand{\arraystretch}{1}
\end{flushleft}
\end{table}

\begin{figure*}[!ht]
\resizebox{\hsize}{!}{\includegraphics{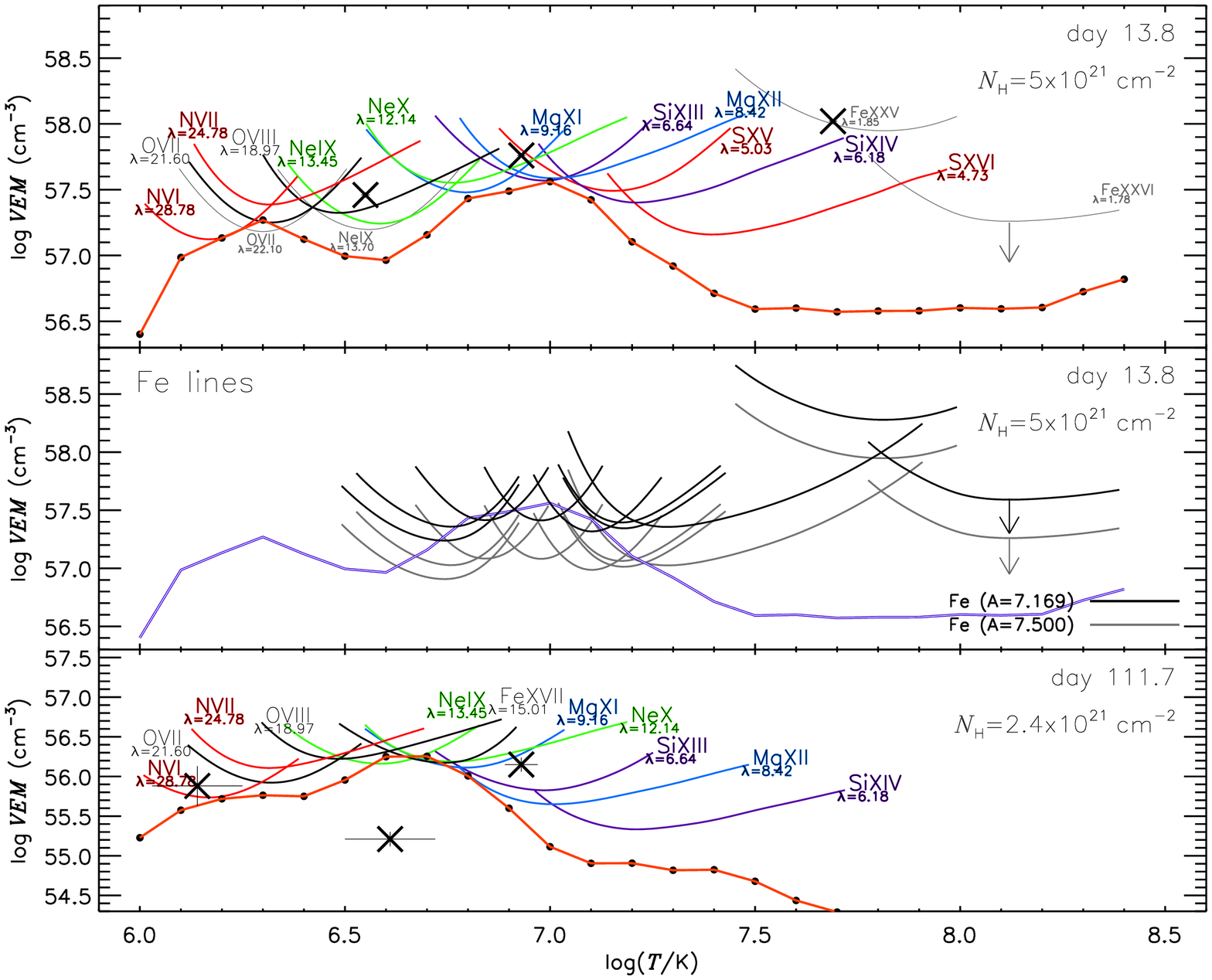}}
\caption{\label{emd}{\bf Upper panel}:
Emission measure loci $L_\lambda(T)$, calculated for each line using
Eq.~\ref{locieq} with line fluxes measured on day 13.8, and
line emissivities assuming rescaled solar abundances
\citep{grev} using the scaling factors
listed in the top part of Table~\ref{abutab}.
The black bullets, connected by a thick solid line
indicate the mean emission measure distribution (EMD)
yielding the best reproduction of the measured line fluxes.
The loci for some lines that are not used to optimize the
EMD are shown with light gray. The black X symbols indicate
the results from 3-$T$ APEC models listed in Table~\ref{xspec}
(see \S\ref{xspecsect} and discussion in \S\ref{cmpmodels}).
{\bf Middle panel}: Emission measure loci only for
Fe lines (gray: assuming solar abundances, black:
corrected by factor 0.46 times solar) and the best-fit
EMD from the top panel in purple.
The bulk of the lines demands a reduction of the Fe abundance,
but the Fe\,{\sc xxv} locus (top right curve) is then too high.
The locus of Fe\,{\sc xxvi} (far right curve) is an upper
limit. The contribution function of H-like Fe\,{\sc xxvi} above
$10^8$\,K is estimated by extrapolating from the data below
$10^8$\,K assuming the same shape as the H-like line of S\,{\sc xvi}.
{\bf Bottom panel}: Same as top panel for line fluxes measured
on day 111.7. The same elemental abundances are used. The black
X symbols indicate the results from APEC models listed in
Table~\ref{xspec}.
}
\end{figure*}

We use the measured line fluxes listed in Table~\ref{elines}
(corrected for absorption; see below) in order to reconstruct
a continuous mean emission measure distribution (EMD) as a
function of temperature, i.e., $VEM(T)$. We assume a constant
electron pressure of $\log(P_{\rm e})=13.0$ (in units
K\,cm$^{-3}$), which is equivalent to a density
$\log(n_{\rm e} [{\rm cm}^{-3}])\lesssim 7$, depending on
temperature. While this assumption is more realistic than
$\log(n_{\rm e})=0$ (as used for the APEC models) this is still
very crude, however, we use only lines that are not
density-sensitive such that the results do not depend on the
assumed pressure.

We concentrate on the two simultaneous observations taken on day
13.8 because the combined \chandra\ and \xmm\ data provide the
largest coverage in lines and most reliable line flux
measurements. Details of the method are described in
\cite{nejor07}. A similar approach has been described by
\cite{ness_vel}. A continuous EMD is a more realistic
representation than multi-temperature models, and it is a
way to overcome
the extreme simplification of assuming an isothermal plasma
\citep[e.g.,][]{bode06}. We use Eq.~\ref{fpred} to compute line
fluxes from a given EMD and compare the predicted fluxes to the
measured fluxes. For Eq.~\ref{gte} we assume the same
ionization balance that \cite{nejor07} used. The elemental
abundances are relative to solar by \cite{grev}.

As a guide to construct a starting EMD we compute the so-called
emission measure loci $L_\lambda(T)$, which are the ratios of
the measured line fluxes, $f_\lambda$, and the line contribution
functions, $G_\lambda(T)$ (Eq.~\ref{gte}), i.e.,
 \begin{equation} \label{locieq}
L_\lambda(T)=\frac{f_\lambda}{G_\lambda(T)}\,.
\end{equation}
 In Fig.~\ref{emd} we show these loci for a set of lines selected
on the grounds that the atomic physics are reliable and that
the measured line fluxes (corrected for $N_{\rm H}=5\times
10^{21}$\,cm$^{-2}$ for day 13.8 and $N_{\rm H}=2.4\times
10^{21}$\,cm$^{-2}$ for day 111.7) are either not blended
with other lines or
are easy to deblend (see comments in Table~\ref{elines}). A few
other lines are shown for comparison in light gray with the label
at their minima. Since the line contribution functions
$G_\lambda(T)$ scale with the elemental abundances (see
Eq.~\ref{gte}), a reduction in abundances leads to an increase
of $L_\lambda(T)$ at all temperatures and vice versa.
Because most Fe lines are quite weak, difficult to
measure, and are subject to less certain atomic physics
because of the complex ion structures, we exclude all Fe
lines from constraining the mean EMD.

 To find a model that reproduces the selected line
fluxes we construct an initial EMD by eye. We
start with the envelope curve below the minima of all
$L_\lambda(T)$ curves and make
successive changes to the EMD and to the elemental abundances
until the predicted line fluxes agree qualitatively with the
measured values. We adjust only the abundances of elements
that produce strong lines, except for oxygen. The oxygen line
fluxes pose a constraint on the normalization, and all
abundances are thus relative to oxygen. Since the line
contribution functions are
broader than 0.2 dex, no narrow features in the temperature
distribution can uniquely be resolved, and we thus allow no
features in the EMD that are narrower than the line contribution
functions. We stress that with our approach we are determining
only one possible representation of the true nature of the shocked
plasma since Eq.~\ref{fpred} represents a Fredholm integral
equation which is not uniquely solvable.
However, \cite{nejor07} pointed out that the determination of
elemental abundances seems fairly robust against the precise
form of the assumed mean emission measure distribution.

 Once a reasonable model is found we fine-tune the model
by iteration of the mean EMD, optimizing the predicted
line fluxes using the method described by \cite{nejor07}.
Based on the ratios of measured to predicted line fluxes
for the best-fit models, we modify the abundances of
elements where systematic discrepancies can be identified
and repeat the fine-tuning of the EMD. In this way we
consecutively approach a good representation of all lines
included in the fit as well as some other lines that are not
included.

 The final model is indicated with the thick solid line in
the top panel of Fig.~\ref{emd}. The best fit yields two
peaks at $\sim 10^7$\,K and $\sim 2\times 10^6$\,K.
The high-temperature regime is poorly determined because the
only lines formed at temperatures above $\log T=7.5$ are
those of Fe\,{\sc xxv} and Fe\,{\sc xxvi}. For the latter
line we only have an upper limit to the flux. 
The use of these lines is limited by the unknown Fe
abundance at this stage.

 In the middle panel of Fig.~\ref{emd} we show
emission measure loci of ten Fe lines measured from the
MEG and HEG spectra taken on day 13.8. The grey curves are
the loci calculated with solar Fe abundance, and all loci
around $\log T\sim 7$ need to be raised (yielding a
reduction of the Fe abundance) in order to be
consistent with the mean EMD derived from the other lines.
If the Fe abundance is reduced by a factor
0.5 (black loci), the reproduction of the Fe lines improves
significantly. Only, the Fe\,{\sc xxv} line is not reproduced,
yielding an underprediction by a factor of 30.

The excessively high Fe\,{\sc xxv} flux in combination with
the non-detection of the Fe\,{\sc xxvi} line is difficult to
explain. While the underestimated flux for Fe\,{\sc xxv}
could be fixed with more emission measure at high
temperatures, such a modification demands a detectable flux
of the Fe\,{\sc xxvi} line. When increasing the emission
measure only at temperatures where the Fe\,{\sc xxv} lines
are formed, the S\,{\sc xvi} line is significantly
overpredicted. We are confident that the
Fe\,{\sc xxv} emissivity function is not underestimated,
as the two atomic data bases APEC and CHIANTI give
consistent emissivities and it seems unlikely to us that both
databases would give the wrong emissivities for such a
relatively simple (He-like) ion. 

Another possibility is that the underlying assumptions
for the calculation of the predicted line fluxes for
Fe\,{\sc xxv} are incorrect. Our assumption of constant pressure
$\log P=13$ implies that Fe\,{\sc xxv} is formed in an
environment of $\log n_e\sim 5-6$, thus a rather low density.
However, in order to make a significant difference in
the predicted Fe\,{\sc xxv} lines the density would
have to be in excess of $10^{16}$\,cm$^{-3}$, and we
reject this possibility (see also lower part of
Table~\ref{lytemps}). The Fe\,{\sc xxv}
flux could be enhanced by resonant scattering into the
line of sight if the plasma is not optically thin. This
would only affect the resonance lines, and the Fe\,{\sc
xxvi} line would also have to be enhanced but it is not detected.
While the emissivities used to compute the Fe\,{\sc xxv}
locus curve uses the combined emissivities from the resonance,
intercombination, and forbidden lines, contributions from
unresolved satellite lines are neglected. These can dominate
the Fe\,{\sc xxv} complex at temperatures below the peak
formation temperature (i.e., at $\log T<7.8$;
\citealt{fe25_satellites}). Lastly, a number of
non-equilibrium processes have significant effects
on the Fe\,{\sc xxv} complex (see \citealt{fe25}).
For example, recombination into excited states would 
enhance the Fe\,{\sc xxv} lines at the expense of
Fe\,{\sc xxvi} and could explain the unusually high locus
of the Fe\,{\sc xxv} lines. While significant effects of
recombination of Fe\,{\sc xxvi} into Fe\,{\sc xxv}
violates the underlying assumption of collisional
equilibrium, this does not necessarily imply that the
other lines are also affected. Recombination affects
only the ionization stages whose ionization energy is
higher than the kinetic energy of the hottest plasma
component. The energy required to ionize Fe\,{\sc
xxv} into Fe\,{\sc xxvi} is 8.8\,keV (equivalent to
$10^8$\,K), and that is higher than the hottest
plasma component found from the APEC models
of $\log T=7.69$ (Table~\ref{xspec}). Meanwhile,
the ionization temperature of Fe\,{\sc xxiv} into
Fe\,{\sc xxv} is only 2.04\,keV ($10^{7.38}$\,K)
which is clearly lower than the hottest plasma
component, and Fe\,{\sc xxiv} and Fe\,{\sc xxv}
can be considered to be in equilibrium. Also,
S\,{\sc xvi} and S\,{\sc xv} are in equilibrium, since
the ionization energy is 3.224\,keV ($=10^{7.57}$\,K).
We therefore conclude that only Fe\,{\sc xxv} and Fe\,{\sc
xxvi} might not be in equilibrium while all other lines
are, and our underlying assumptions are valid for these
lines.

In Table~\ref{fluxpred} we list the measured and predicted
line fluxes for our best EMD for day 13.8 (Fig.~\ref{emd}),
giving element, rest wavelength,
measured fluxes (after deblending and correction for
$N_{\rm H}=5\times 10^{21}$\,cm$^{-2}$), and ratios of measurements
and predictions. We assume a value of $N_{\rm H}$ that is lower
than that found from the APEC models (Table~\ref{xspec})
because we are unable to find an EMD model with that value
that gives such good reproduction of all line fluxes. Also,
with lower values of $N_{\rm H}$ we are having difficulties
to find a good EMD model, and values as low as the interstellar
value of $N_{\rm H}=2.4\times10^{21}$\,cm$^{-2}$ can be
excluded. We note that the EMD modeling is not an ideal way
to determine $N_{\rm H}$, and we refrain from determining
a confidence range for $N_{\rm H}$ but consider it an
uninteresting parameter, thus concentrating on the elemental
abundances. We note that the chosen value of $N_{\rm H}$ is
consistent with the total column density found by \cite{bode06}.

 In Table~\ref{abutab} we give the correction factors
applied to the abundances. The EMD is scaled to reproduce the
oxygen lines assuming solar O abundance (correction is 1.0),
and the correction factors are thus equivalent to the
respective abundances relative to solar. We also
give the logarithmic abundances in the standard notation
and list abundances relative to Fe (computed from the respective
values relative to oxygen) in the bottom of Table~\ref{abutab}.
We determine the
uncertainties of abundances by stepping through a grid of
values, each time readjusting the EMD and computing a value
of $\chi^2$ from the measured fluxes with their uncertainties
for the selected lines. The listed uncertainties are derived
from increases of $\chi^2$ by one, which in the case of a
1-parameter model would be the 1-$\sigma$ uncertainties. We
note that our model is not a parameterized model.

 In the absence of any hydrogen lines, absolute abundances
can only be determined via the strength of the
continuum relative to the lines (see above). Since the
mean EMD is not well constrained at high temperatures, the shape
of a continuum model predicted by the EMD disagrees with the
observed spectrum. It is not possible to adjust the EMD without
conflicts with some of the emission lines, and we thus refrain
from determining absolute abundances with this method.

 We apply the same approach using the line fluxes from
other observations listed in Table~\ref{elines}, and show the
results in the bottom panel of Fig.~\ref{emd} for day 111.7.
All emission measure values are significantly lower, and
there is no indication for plasma hotter than $\log T=7$.
The abundances are given in the last column of
Table~\ref{fluxpred}, and they are all consistent with
those found on day 13.8.
This conclusion also holds for the other datasets,
although it is more difficult to distinguish between different
EMD models. The reasons are lack of lines
formed at low temperatures for the observations taken on
days 26.1, 39.7, 54.0, and 66.9 and lack of lines formed at
high temperatures for the observations taken after day 111.7.
Also, some lines are blended with other nearby lines which
can be disentangled with the \chandra\ HETGS, but not
with the \xmm\ RGS and \chandra\ LETGS.

\begin{figure*}[!ht]
\resizebox{\hsize}{!}{\includegraphics{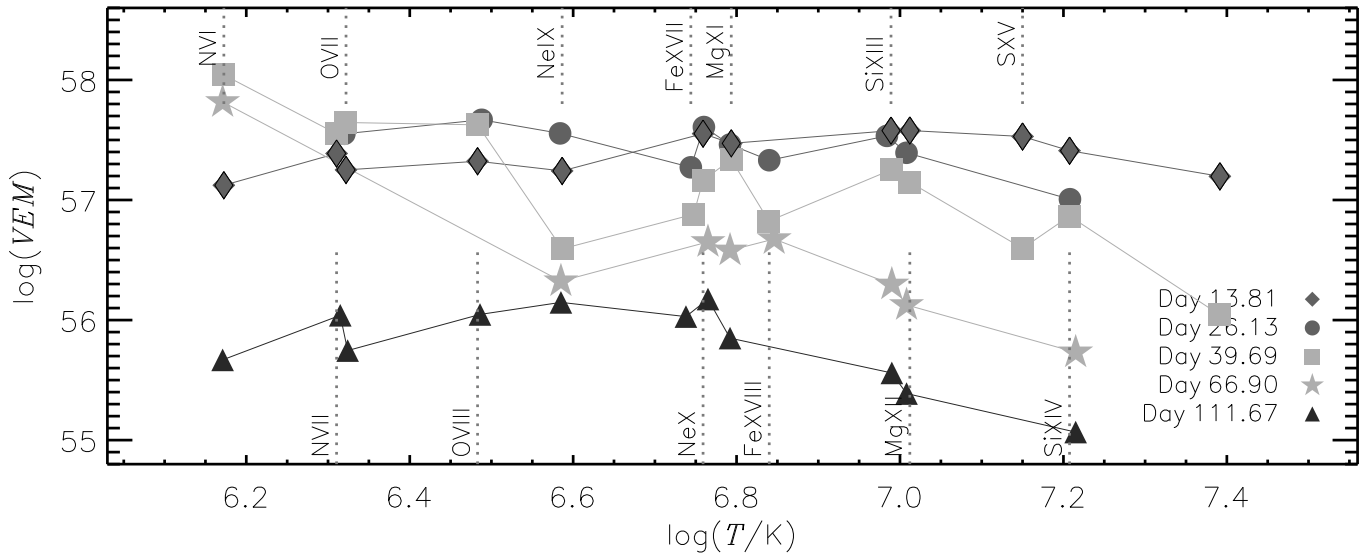}}
\caption{\label{loci}Volume emission measures ($VEM$) at the peak
formation temperature for selected strong lines, assuming
the abundances listed in Table~\ref{abutab}. The connecting
lines guide the eye, but are not EMD models. The plot symbols
indicate the time of observation after outburst according to
the key in the bottom right legend. We use dark gray for
observations taken before the SSS appeared, light gray for
observations during the SSS phase and black for
after the SSS spectrum had disappeared.}
\end{figure*}

In Fig.~\ref{loci} we show the minima of the emission
measure loci for all observed line fluxes using different
plot symbols for each observation as explained in the
legend. To guide the eye we connect the datapoints belonging
to the same observations with solid and dotted lines in order
that the evolution of the temperature structure can be
identified. A similar plot has been presented by
\cite{schoenrich07} who assumed the same value of
$N_{\rm H}=2.4\times10^{21}$\,cm$^{-2}$ for all observations
and solar abundances. They found significantly higher
loci for the N and O lines for days 39.7 and 66.9 compared
to all other observations. Since during the SSS phase these
lines appeared on top of the SSS continuum, they concluded
that these lines are formed within the outflow or are at
least affected by the SSS radiation. We now use the new
abundances from Table~\ref{abutab} and
$N_{\rm H}=5\times10^{21}$\,cm$^{-2}$ for days 13.8 and 26.1
and $N_{\rm H}=2.4\times10^{21}$\,cm$^{-2}$ for the rest
(see Table~\ref{xspec}). The higher value of $N_{\rm H}$
for days 13.8 and 26.1 leads to higher loci of the
low-temperature lines of O and N which are now consistent
with those measured for day 39.7 and 66.9. In order to
attribute these lines to either the outflow or the shock
therefore requires accurate knowledge of the value of
$N_{\rm H}$. But even with our improved measurements of
$N_{\rm H}$ the situation remains ambiguous because,
if the O and N lines observed on top of the SSS continuum
are formed somewhere within the outflow, they might be
subject to higher values of $N_{\rm H}$ as is suggestive
from the shock plasma. This would increase the
discrepancies again, and with these uncertainties, it may
never be possible to decide whether these emission lines
are formed in the outflow or in the shock.

 At the other end of the temperature distribution, at
$\log T>6.8$, a steady decrease of emission measure can be
identified. The slope of the temperature distribution 
toward the highest temperatures seems to become steeper,
indicating that the hot plasma is cooling rapidly, while
below $\log T=6.8$ the line emission measures also decrease,
but the slope remains about the same, and the cool component
thus cools at a slower rate.


\subsection{Comparison of models}
\label{cmpmodels}

While the APEC models introduced in \S\ref{xspecsect} include all
available atomic information, the line-based approach used in
\S\ref{lfluxes} has the advantage that only the most reliable
information is
selected. Less certain lines (e.g., lines with transitions
involving higher principal quantum numbers) are discarded. In
\S\ref{lfluxes} the crude assumption of constant pressure
is sufficient because only lines that are not density-sensitive
are selected. The APEC model makes the even cruder assumption
of a low-density plasma \citep[$\log n_e=0$:][]{smith01},
while density-sensitive
lines are not excluded. However, most density-sensitive lines
are weak. Another difference is the assessment of
the goodness of a model. In our line-based approach the fitting
yields best reproduction of measured line fluxes (taking line
broadening into account), while with {\sc xspec} the count rate
in each spectral bin has to be reproduced. In cases where lines
are present in the observed spectra that are missing in the atomic
database, we can ignore them with the line-based approach, while
with {\sc xspec} the existing lines in the atomic database can be
used to force an acceptable fit of these spectral regions.
Furthermore, the individual emission lines contribute only to a few
spectral bins such that negligence of emission lines is
badly penalized compared to negligence of the continuum.
On the other hand, the APEC models are much better suited to
assess the hottest temperature component via the continuum, and
the corresponding shock velocities can only be derived from the
APEC models (see Fig.~\ref{evol_v}).
For more discussion on line-based and global fitting approaches
we refer to \cite{cospar04}.
We thus need both approaches for robust conclusions.\\

 In the top panel of Fig.~\ref{emd} we include the
temperatures and volume emission measures of the three
components derived from the 3-$T$ APEC models fitted to
the spectra of day 13.8 for comparison with the mean EMD.
All three components yield higher values of emission measure
than the continuous EMD, because all emission from the
smooth distribution of the EMD model is concentrated in
only three isothermal components. The emission measure of the
hottest component of the APEC model is much higher than in
the EMD model. Since not enough emission lines are
formed above $10^{7.5}$\,K to constrain the EMD,
the line-based approach is clearly inferior in this
temperature regime, and the hottest component
is driven by the continuum. We further find that only
the lines that are formed at high temperatures are
affected by recombination (see \S\ref{lfluxes}).
The second temperature component of the APEC model
coincides with the temperature of the hotter peak of the
mean EMD. The
third component is closer to the minimum between the two
peaks of the mean EMD (see Fig.~\ref{emd}). The higher
temperature in the APEC model could explain why the N
abundance is higher than that derived from the EMD
modeling. Since each temperature component is isothermal,
the N lines are not as efficiently produced by the cool
component, which has to be compensated by a higher N
abundance in order to fit the N lines. We also note that
the N\,{\sc vi} line at 28.78\,\AA, which is clearly
detected (see Table~\ref{elines}), is completely ignored
by the APEC models but represents an important constraint
on the EMD model. Finally, in the {\sc xspec} fits we
could not account for line broadening (see
\S\ref{xspecsect}), while we used the line fluxes
integrated over the entire profile for the EMD reconstruction
method. We therefore regard the abundances of O and N
derived from the EMD model as more reliable than the
values derived from the APEC model.

For all models, we give only statistical uncertainties.
Uncertainties from the atomic physics are not included
and are a source of additional systematic uncertainty
(\S\ref{syserr}). Since many lines with poorly-known
atomic physics are included in the APEC models the
systematic uncertainties of the APEC models are higher
than those from the EMD models.

For comparison of the elemental abundances derived from
the two approaches, we rescale those obtained in
\S\ref{xspecsect} because the reference abundances used
in the APEC models are those by \cite{agrev89}, while
for the EMD reconstruction method we have used
those by \cite{grev}. We rescale by a factor
$(N_{\rm X}/N_{\rm O})_{\rm grev}/(N_{\rm
X}/N_{\rm O})_{\rm and}$, where the subscripts
'grev' and 'and' denote the abundance ratios from \cite{grev}
and \cite{agrev89}, respectively. The correction
factors are 0.933, 1.23, 1.259, 1.259, 1.66, and 0.85
for N, Ne, Mg, Si, S, and Fe, respectively. In Table~\ref{abutab}
all derived abundances are listed for comparison. While
the abundance ratios relative to O are discrepant, those
relative to Fe agree much better, except for N/Fe.
We attribute these differences to the low formation
temperatures of the N and O lines. We attribute these
differences to the less certain N and O abundances
derived from the APEC models that underestimate the
amount of cool plasma (see above). We expect overestimated
N and O abundances and possibly also Ne in the APEC model
which explains that all abundances relative to O are lower
in the APEC model compared to the EMD model. Since the N
lines are formed at lower temperatures than the O lines,
the N abundance is affected to a higher degree. The Fe
lines are formed over a large range of temperatures and are
therefore not as strongly affected, leading to the overall
better agreement of all ratios relative to Fe. For the
observation taken on day 111.7 only the EMD modeling yields
some constraints on the elemental abundances, which
demonstrates the strength of this approach over the
spectral fitting.

\subsection{Uncertainties}
\label{syserr}

 While the results from \S\ref{obssect} are directly based on
the observations, all results from this section, \S\ref{anal},
depend on model assumptions which are described in
\S\ref{methods}. All error estimates given in this paper
are statistical 1-$\sigma$ uncertainties which give the 68.3-per
cent probability that fitting the same model to a new observation
with the same instrumental setup results in parameters within
the given uncertainty ranges. They thus only describe the
precision of our measurements, but not the accuracy (sum of
statistical and systematic uncertainties) which depends on
the calibration of the observations but also on the choice of a
model. For comparisons of measurements taken with the same
instrument (given in Table~\ref{tab1}), the systematic
calibration uncertainties can
be neglected but have to be kept in mind for absolute
numbers and comparisons with different instruments (e.g.
the flux evolution shown in Fig.~\ref{evol}). For line
profiles (Fig.~\ref{v13}) and line ratios (Fig.~\ref{lrats}),
the cross-instrument calibration uncertainties are
negligible.

For the flux measurements presented in Fig.~\ref{evol},
systematic uncertainties arise from the choice of band width
which excludes a fraction of the total X-ray emission.
Owing to the presence of the SSS spectrum between days 39.7
and 66.9, the contribution from emission between
11 and 38\,\AA\ to the shock emission can not be
determined, yet the fraction of soft emission may be higher
compared to before day 39.7. Attempts to determine this
contribution from the {\sc xspec} models failed because
we have no constraints from observations because the
much stronger SSS emission dominates at long wavelengths.
We estimate that the systematic
uncertainties on the \swift\ light curve shown in
Fig.~\ref{evol} are small,
because the spectral shape hardly changes after day 100
(see Table~\ref{xspec}), and the X-ray
flux thus scales directly with the observed count rate.
We note that systematic uncertainties from direct rescaling
are smaller than the method of flux determination via
model fitting to each individual \swift\ spectrum.

For the line shift measurements presented in Fig.~\ref{v13}
and \S\ref{lprofiles}, systematic uncertainties can arise
from line blends and background noise, which affects the
weaker lines more than stronger lines. For the
measurement of line fluxes we have not accounted for
uncertainties from choice of a source background
(continuum) underneath the lines and from the line
widths. \cite{nejor07} found that the uncertainties
in the line widths have less effect than the choice of
continuum. For the line ratios
presented in Fig.~\ref{lrats}, additional systematic
uncertainties from the continuum are less than
5\,per cent if the continuum is assumed to be
uncertain at the 20\,per cent level.

 For the model fits, the main sources of systematic
uncertainty are the model assumptions and uncertainties
in the atomic data. Both are difficult to quantify. The
hottest temperature components of the 3-$T$ fits presented
in Table~\ref{xspec} are
driven by the continuum. The Fe\,{\sc xxv} and
Fe\,{\sc xxvi} lines are the only strong lines formed
at temperatures higher than $\sim 10^7$\,K, and their
absence in a model poses little resistance against an
overestimated continuum temperature. For observations
with weak continuum, arbitrarily high temperatures
can thus result, and we give higher confidence to
the temperatures of the hottest component determined
from observations before day 39.7 (see Table~\ref{xspec}).
In view of these large systematic uncertainties of
the later observations, the trends of shock velocities
plotted in Fig.~\ref{evol_v} are only reliable for the
early observations. Since the apparent $t^{-1/3}$ trend
that is suggestive from the values derived for
day 39.7 and 111.7 is based on the higher temperature
estimates compared to the expected $t^{-1/2}$ trend,
the conclusion of adiabatic cooling instead of
the theoretically expected radiative cooling has
to be treated with caution.

 The uncertainties in
elemental abundances resulting from the emission
measure modeling (Table~\ref{abutab}) are extremely
difficult to quantify. They mainly depend on the
uncertainties in the line flux measurements and
the uncertainties in the determination of the
mean EMD. \cite{nejor07}
found from comparison to independent analyses
of the same spectrum by \cite{jsf03} that the
abundances are quite robust against changes in
the assumed mean EMD. Since the abundance
determination is based on strong lines originating
from few-electron ions (thus well-known atomic physics),
we are confident that the given uncertainties
that are based on the measurement uncertainties
of line fluxes are realistic.

\section{Discussion}
\label{disc}

The X-ray grating spectra give the deepest insight
into the properties of the shocked plasma. With 12 grating
observations we can follow the changes in the X-ray flux and
plasma temperature, and we have determined the elemental abundances
from the emission lines. The X-ray flux can directly be
integrated over the spectrum without the need of a model
and can thus be considered an observed quantity. We are
further able to
determine temperatures independently from line flux ratios
and from spectral models, and we are able to detect different
temperature components. The ability to measure line fluxes
also enables us to pursue two independent approaches to determine
the elemental abundances, yielding robust results.
We confirm that the underlying assumptions are generally
satisfied, but we find evidence for recombination in the
Fe\,{\sc xxv} He-like triplet lines and the Fe\,{\sc xxvi}
lines. Since these
are the only ions whose ionization energy exceeds the
kinetic energy of the hottest plasma component, we regard
the assumptions discussed in \S\ref{anal} valid for all other
lines.

We observe a power-law decay in X-ray flux, and the power-law
index changes from $\alpha=-5/3$ before day $\sim
70$ to $\alpha=-8/3$ during the later evolution (see
Fig.~\ref{evol}). While the $t^{-5/3}$ decay is predicted for
a radiatively cooling shock traversing an $r^{-2}$ density 
distribution, the change to a t$^{-8/3}$ decay at late times
cannot be simply explained. However, note that \cite{vaytet07}
found that in some circumstances the forward shock could
decelerate faster than the standard cooled momentum-conserving
models predict. In addition, it is interesting to note that a
steeper decay may 
result from the breakout of the forward shock into a lower density 
environment when it reaches the edge of the red giant wind re-established 
in the 21 year interval between explosions. Indeed, simple calculations 
suggest that this breakout may happen around the time of the apparent 
break in decay curves (see, e.g., \citealt{mason87}; \citealt{obrien92}).
One possible problem with this conclusion however is that the 
immediate post-shock temperature increases, rapid
adiabatic expansion of the emitting material will likely lead to a drop in
the temperature derived from fits to the X-ray spectra, which does not seem
to be the case. Furthermore,
\cite{bode_keele} conclude from the apparent early onset 
of the adiabatic expansion phase in the forward shock that shock breakout 
will occur much later than realized from X-ray observations that in 1985 
started only at 55 days post-outburst.

 The two peaks
in the emission measure distribution (see Fig.~\ref{emd})
could represent the forward and reverse shocks, where the
cooler component would be the reverse shock. But according to
models by \cite{bodekahn85}, the reverse shock may not be a
significant source of X-ray emission at this time. Alternatively,
the two temperature regimes could relate to forward shocks
propagating into two distinct circumstellar density regimes,
e.g., denser equatorial regions and less dense polar regions
as suggested by VLBA/EVN and HST observations
\citep{bode07,obrien06}. More realistic hydrodynamic modelling
is needed, and this is underway \citep[see][]{vaytet07}.

 In several Classical Novae, e.g., V382\,Vel, X-ray spectra
similar to those of the late evolution in RS\,Oph have been
obtained after the nova had turned off and were attributed
to the ejecta that are radiatively cooling \citep{ness_vel}.
It is thus possible that the cooler component observed after
day 70 originates from the expanding ejecta, while the early
hot and cool components represent the shock emission.

\subsection{Elemental Abundances}

 For our EMD models and multi-temperature fits, we have
assumed uniform abundances throughout the X-ray emitting
plasma. If the relative contributions from each component
to the observed X-ray emission change with time and their
compositions are significantly different, then in principle,
changes in observed abundances derived from different
observations are possible. However, the relative abundances
obtained for days 13.8 and 111.7 do not differ significantly,
and we conclude that no signs of changes in the composition
are detectable. We therefore focus on our
measurements for the best dataset taken on day 13.8.
There are several issues for which the accurate determination
of elemental abundances is of interest.
The composition of the WD can be used as an argument
whether or not RS\,Oph will explode as a canonical SN\,Ia. The
WD may be near the Chandrasekhar limit, because of the short
recurrence time scale, but if the WD consists of too many
heavy nuclei (Mg, Ne rather than C, O), the amount of
available nuclear binding energy is not sufficient to totally
disrupt the WD. Note that it is the explosions of CO
WDs that best fit the observations of SN\,Ia
\citep{leibundgut2000,hillnie2000}.

Since hydrogen is fully ionized in X-ray emitting plasma
no H lines can be observed, and we cannot determine
absolute abundances. In both of our approaches the attempt
to determine absolute abundances from the strength of the
continuum relative to the lines leads to ambiguous results,
and we thus focus on relative abundances.

\begin{figure}[!ht]
\resizebox{\hsize}{!}{\includegraphics{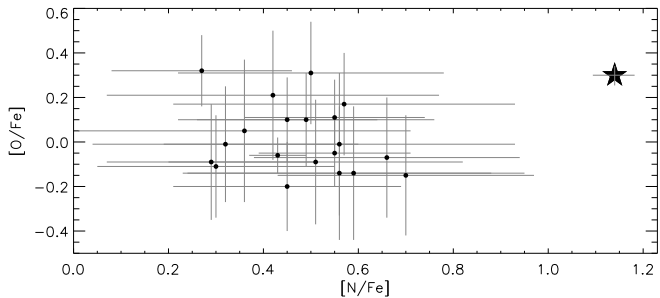}}
\caption{\label{cno_abus}Photospheric abundance ratios
of 20 M giants given by \cite{smithlambert85,smithlambert86}.
The star symbol is our measurement for RS\,Oph as listed
in Table~\ref{abutab}. Uncertainties are given by
gray error bars.}
\end{figure}

\noindent {\em CNO-cycled material}\\
There is no doubt, from both of our approaches, that
nitrogen is overabundant. This points to CNO-cycled
material, which could have been produced during the
outburst. However, the companion is a red giant and is
as such likely overabundant in nitrogen as well
\citep[see Fig.~\ref{cno_abus}
and][]{smithlambert85,smithlambert86,cno}.
The accreted material on the WD is thus already N-enhanced
\citep{snijders87}, as is the wind into which the ejecta run.
In addition, nuclear burning during the outburst can contribute
with more CNO-cycled material. Furthermore,
N-enhanced material from previous outbursts may also
contribute to the X-ray emitting plasma. Note that
\cite{walder08} have determined from theoretical models
that the ejecta are dominated by the RG material.
In Fig.~\ref{cno_abus} we show the N abundance measurements
of 20 M giants presented by \cite{smithlambert85,smithlambert86}
versus their O measurements. The star symbol indicates
our result for RS\,Oph, and it is clearly higher than any of the
M giants, while the O abundance is consistent with most
of the M stars. The comparison with other M giants thus
suggests that CNO-processed material may have been added by the
outburst. In support of this, the N/O abundance ratio derived
from our two approaches is roughly similar to the value
determined by \cite{schoenrich07} ($[$N/O$]=0.7$), who
determined the ratio from the line column densities in the
SSS spectra. Our result is also consistent with
the abundance measurements by \cite{contini95} which yield
an abundance ratio of $[$N/Fe$]=1.18$.
 The photospheric abundance measurements of the RS\,Oph
secondary presented by \cite{pavlenko} yield $[$N/Fe$]=0.9$ 
which can be compared to our abundance ratio $[$N/Fe$]=1.14$.
This is lower than either our value or that determined by
\cite{schoenrich07} ($[$N/Fe$]=1.0$ under the assumption
of $[$O/Fe$]=0.3$), suggesting that the outburst has
added about 20-40\% of nitrogen into the X-ray
detected material. We note that $[$N/Fe$]$ as derived by
\cite{pavlenko} is higher than any of the values plotted in
Fig.~\ref{cno_abus}. This indicates that the secondary in the
RS\,Oph system is not a typical M giant which may be caused
by the fact that the RS\,Oph secondary has lost
significant amounts of material from its outer envelope
during accretion to the WD.

If the material was CNO cycled, then we would expect it to
be underabundant in carbon \citep[see, e.g.,][]{cno}, but
all carbon lines are at wavelengths where strong effects by
interstellar and circumstellar absorption dominate. Moreover,
uncertainties in $N_{\rm H}$ propagate to large uncertainties
in any line fluxes or upper limits of the C lines (see
Table~\ref{elines}). For any quantitative assessment of CNO
burning, the carbon abundance is essential
\citep{ghoul}.

\cite{nelson07} discussed the carbon abundances and
obtained an underabundance of C/N by a factor of 0.001 in
order to explain the absence of the C\,{\sc vi} K-shell absorption
edge in the SSS spectra at 25.37\,\AA. This is much lower
than the C/N abundance ratio of 0.05 that one derives from the
$[$C$]=-0.4$ and $[$N$]=+0.9$, estimated by \cite{pavlenko} for
the secondary. Strong carbon lines were also observed
in the IUE spectra by \cite{shore96}. The absence of the
C\,{\sc vi} edge is not sufficient evidence for a C
underabundance because other high-ionization
absorption edges are also absent in the SSS spectra of
RS\,Oph. For example, the ionization edges of O\,{\sc viii}
at 14.23\,\AA, O\,{\sc vii} at 16.77\,\AA, N\,{\sc vii} at
18.50\,\AA, and N\,{\sc vi} at 22.457\,\AA\ are not present
although these elements are not underabundant (see figs.~3 and 4 in
\citealt{ness_rsoph} and figs.~8 and 9 in \citealt{nelson07}).
Since \cite{ness_rsoph} measured considerable blue shifts
in the absorption lines observed during the SSS phase of
RS\,Oph, the shell around the WD must be expanding with
high velocities. While \cite{nelson07} found no blue-shifts
in the same lines, they quote high velocities derived from
line shifts of emission lines of C\,{\sc vi} Ly series
lines, however, without the presence of the Ly$\alpha$
line (Table~\ref{elines}). For an environment with high
expansion velocities, the absorption edges may be washed out by
the expansion. We therefore argue that without measurements
of emission line fluxes of carbon, the C abundance cannot
be determined from the X-ray spectra.
Finally, a C/N abundance of 0.001 is far too low for any
material that has undergone CNO cycle processing. The large
N abundance requires some processing back to C
\citep{starrfield08}.\\

\noindent {\em $\alpha$ elements ($Z<22$)}\\
The abundances of the $\alpha$ elements Ne, Mg, Si, S are
significantly higher than O and Fe using both approaches. We
use O as the reference element for our analysis, and our
Ne/O abundance ratio is consistent the value found from
IR observations by \cite{nye07}. This indicates that
either Ne is overabundant or O is underabundant.
\cite{contini95} found significant underabundance of O/H
and of Ne/H of $\sim 10$\% solar. We also list the ratios
relative to Fe in Table~\ref{abutab}. The abundance
determinations by \cite{contini95} also yield high
abundance ratios of $[$Mg/Fe$]=0.73$ and $[$Si/Fe$]=0.86$,
which are roughly consistent with our results.
Since Fe is neither produced nor destroyed in the outburst,
we argue that all $\alpha$ elements are overabundant
in the X-ray emitting plasma. Since we determined these
values with two independent approaches using spectra with
strong lines of relatively simple ions with well-known atomic data,
we regard these results as reliable.

 There are three possible explanations: (1) the $\alpha$ elements
were produced during the outburst, (2), the composition of
the secondary, which provides the accreted material and
dominates the composition of the stellar wind, is overabundant
in these elements (3) WD core material is enriched in the
$\alpha$ elements, and that material has been mixed into the
outflowing material.

(1) The production of $\alpha$ elements via nuclear burning
requires high temperatures and densities. While the
required temperatures are likely reached during the TNR,
the densities are too low in order to produce significant
amounts of $\alpha$ elements, even if the WD mass is near
the Chandrasekhar mass.

\begin{figure}[!ht]
\resizebox{\hsize}{!}{\includegraphics{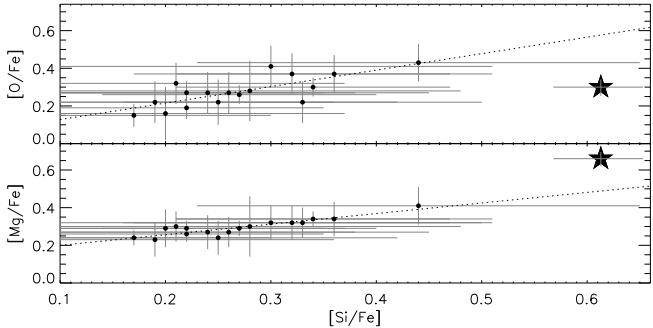}}
\caption{\label{alpha_abus}Photospheric abundance ratios
(relative to solar by \citealt{grev}) of 17 M
giants given by \cite{rich07}. The star symbol is
our measurement for RS\,Oph as listed in Table~\ref{abutab}.
Uncertainties are given by gray error bars.}
\end{figure}

(2) Direct abundance measurements of $\alpha$ elements
for the secondary are not available, and
\cite{pavlenko} only determined Fe, C, and N abundances.
\cite{rich07} measured photospheric
abundances for 17 M giants in the inner bulge of the
galaxy from IR spectroscopy. In Fig.~\ref{alpha_abus}
we show their abundances of $[$Si/Fe$]$ versus
$[$O/Fe$]$ (top) and versus $[$Mg/Fe$]$ (bottom),
relative to
\cite{grev}. All M giants are overabundant in Si,
O and Mg. Stars with a higher Si abundance are also
higher in Mg and O, indicating different concentrations
of the ashes of He shell flashes, probably due to different
stellar ages. Our Si and Mg abundances
for RS\,Oph are significantly higher than any of the M giants
in the sample, but the O abundance is not higher. The
bottom panel of Fig.~\ref{alpha_abus} shows that our
value is significantly higher in both elements than any
of the M giants in the sample,
and the concentration of all $\alpha$-elements
could be higher. Since the outer envelope
of the RS\,Oph secondary has been stripped away and
accreted by the WD, higher abundances of all
elements produced by He-burning can be expected.
The surface composition may resemble that of the inner
regions of normal M giants. Meanwhile, the O abundance could
reflect a balance between a higher O abundance produced by
He burning and a lower O abundance in
material that has a higher concentration of CNO processed material
as evident from
Fig.~\ref{cno_abus}. Our measurements can thus reflect
the composition of the accreted material.

(3) If direct abundance measurements of the $\alpha$ elements
of the secondary are different from our values, then the only
remaining possibility would be that the WD contributes significantly
to the composition of the observed plasma.
The composition of the WD is dominated by elements
that have been produced in the progenitor star.
In order to be observable, WD material has to be
dredged up and mixed into the outflowing material. While
this is possible, no observations have so far
revealed any signatures of WD material in the emitting
regions of RS\,Oph.

\subsection{Is RS\,Oph a SN\,Ia progenitor?}

There is some speculation that
RS\,Oph is a SN\,Ia progenitor. The short recurrence
time scale suggests that the underlying WD is of high
mass, possibly near the Chandrasekhar limit
\citep[e.g.,][]{dobrKen94,shore96,fek00}. \cite{hachisu01}
reported a high-mass WD of $\sim 1.35\,\pm\,0.01$\,M$_\odot$
(for both, high- and low-metallicity models), but their
modelling neglects the environment and is not realistic.
An even higher WD mass was estimated by \cite{sokoloski06},
and \cite{hachisu07} also found that the WD is growing in
mass. According to their estimated growth rate, the
Chandrasekhar limit would be reached in a few times
$10^5$ years.

ONe WDs do not provide
enough nuclear binding energy (see, e.g., total binding
energy calculations by \citealt{gamezo03} or \citealt{calder07}),
and when reaching the Chandrasekhar mass limit, they implode
in a core collapse without an explosion. The explosion models
of SN\,Ia predict that the WD has to be a CO WD for the supernova
to be a canonical Ia explosion. However, the WD composition is
not unambiguously determined.

A strong argument against RS\,Oph being a SN Ia progenitor is
the large amount of hydrogen and helium in the system
\citep[e.g.][]{nye06} that has to be removed before the WD reaches
the Chandrasekhar limit in order for the SN Ia outburst to be
consistent with observations of hydrogen-deficiency
\citep{filippenko97}.
While the secondary loses hydrogen during the evolution of
accretion and repeated outbursts, it would have to be considered
too much of a coincidence if all the hydrogen is consumed in
nova outbursts at the same time as the Chandrasekhar limit is
reached. Considering these points, we believe that it is
unlikely that RS\,Oph is a SN\,Ia progenitor.

\section{Summary and Conclusions}
\label{concl}

 The high-resolution X-ray spectra of the 2006 outburst of
RS\,Oph provide unique insights into the properties and
evolution of the outburst. The properties are best
determined from the dataset taken on day 13.8.
At this time of the evolution the nova was brightest
in hard X-rays, and we have the best coverage with almost
simultaneous \chandra\ and \xmm\ observations. From these
observations we have derived the elemental abundances with
an overabundance of N, Ne, Mg, Si, and S, relative to Fe,
indicating that this material has undergone CNO burning
and He burning, respectively.
Both processes can occur in the RG companion star, and our
observations could reflect the donor material. CNO
burning also occurs during the outburst, and by
comparison of our results with direct measurements of
the N abundance for the secondary by \cite{pavlenko}, we
estimate that about 20-40\% of the nitrogen could have
been produced during the outburst. Meanwhile, no $\alpha$
elements can be produced during the outburst, but the
underlying WD might be enriched in $\alpha$ elements.

We find Mg and Si significantly higher than in any M giant
in the sample by \cite{rich07} which can either be explained
by the RG being different from other M giants, or WD material
has been dredged up during the outburst and mixed into the
ejecta. Similarly high values have been found by
\cite{contini95}. Since the secondary has likely been stripped off
it's outer layers by mass loss onto the WD which has then
been ejected during the outburst, the composition of the
ejecta may be sampling deeper layers of the M giant
companion as compared to the photospheric composition
of normal M giants.
In order to determine whether WD material is
observed, direct measurements of the Mg and Si abundance
in the photosphere of RG are needed for comparison.

 We also determined the evolution of the temperature
structure. The early observations show both a hot and
a relatively cooler plasma component, while the later
observations only
display a cool component. Both components decay with
time. Before day $\sim 70$, while the plasma is dominated
by the hot component, the decay rate is slower than later
in the evolution. The early decay rate is consistent
with a radiatively cooling plasma while the later evolution
can be explained by the expansion of the ejecta. The
temperature evolution derived from models are consistent
with radiative cooling of the hot component.

While it has been suggested that the WD is close to
the Chandrasekhar limit, we note that there is currently
too much hydrogen in the system for a supernova explosion
to satisfy the spectroscopic features of a Ia explosion
\citep{filippenko97}. The hydrogen and helium will have to
be removed before the Chandrasekhar
limit is reached. Further, the underlying WD has to be a
CO WD, but the composition of the WD is difficult to determine
since WD material has to be dredged up and ejected. Our
abundance analysis suggests that this may be
the case, but the composition of the accreted material is
not well-enough known for solid conclusions at this time.

\acknowledgments

J.-U.N. gratefully acknowledges support provided by NASA through \chandra\ Postdoctoral
Fellowship grant PF5-60039 awarded by the \chandra\ X-ray Center, which is operated by
the Smithsonian Astrophysical Observatory for NASA under contract NAS8-03060.
S.S. received partial support from NSF and NASA grants to ASU.
J.P.O. and K.L.P. acknowledge support from STFC.
We thank Dr. V. Kashyap for technical assistance using the
PINTofAle tools.
We thank Dr. Y.V. Pavlenko for sharing the results of his abundance
analyses of the RS\,Oph secondary with us.
We are grateful to Harvey Tananbaum and the Chandra
Observatory for a generous allotment of Directors Discretionary
Time to observe this unique outburst.
Some of the observations have been obtained with XMM-Newton, an ESA science
mission with instruments and contributions directly funded by ESA Member States and NASA.
CHIANTI is a collaborative project involving the NRL (USA), RAL (UK), MSSL (UK), the Universities of Florence (Italy) and Cambridge (UK), and George Mason University (USA)

\bibliographystyle{apj}
\bibliography{cn,astron,jn,rsoph}

\appendix
\subsection{Statistics}
\label{appendix}

 Some of the spectra are sufficiently bright for standard
$\chi^2$ minimization, while others are extremely faint, and
low-count statistics, as recommended by \cite{cash79}, have to
be applied. According to Wilks theorem \citep{wilks38}, the
Maximum Likelihood (ML) technique converges to the $\chi^2$
statistics in the high-count limit, and we thus apply the ML
technique to all observations. The likelihood is defined as
\begin{equation}
\label{like}
{\cal L}=-2 \sum_{i=1}^{N}(n_i \ln m_i  - m_i)
\end{equation}
with $n_i$ being an observation and $m_i$ a model,
both defined on the same grid of $N$ bins.
According to Wilks theorem ${\cal L}$
serves as a goodness criterion in an equivalent way as
$\chi^2$, but it is derived from the
Poissonian probability distribution function (PDF) rather
than a Gaussian PDF. In order to conserve the Poissonian
nature of the data, the instrumental background must not be
subtracted. Instead, the model has to be added to the background
for comparison with the raw data, because the sum
of two Poissonian statistics is
Poissonian, while the difference is not \citep{cash79}.\\

 To assess the statistical uncertainty ranges of each
parameter in a multi-parameter model we compute the 
curvatures of the likelihood at
the respective best-fit values from the
second derivatives. In order to assess the correlated
uncertainties, we also compute mixed derivatives, thus the
full Hessian matrix which, for $n$ free parameters, is
an $n\times n$ matrix with the partial second derivatives of
the likelihood curve, ${\cal L}={\cal L}(A_i, A_j)$,
\begin{equation}
\label{hesse}
 H_{i,j}= \frac{\partial {\cal L}^2}{\partial A_i\partial A_j}
\end{equation}
where 
$A_i$ and $A_j$ represent two free parameters with $i$
and $j$ ranging from 1 to $n$. We determine Eigenvectors,
$EV_i(A_j)$ and Eigenvalues, $ev(A_j)$, and compute the
correlated uncertainties from
\begin{equation}
\label{correrr}
 \Delta A_j^2=\Delta{\cal L}\sum_{i=1}^3\frac{EV_i(A_j)^2}{ev(A_j)}
\end{equation}
with $\Delta{\cal L}=2.3$ and $\Delta{\cal L}=3.53$ for
1-$\sigma$ uncertainties in the cases of two and three free
parameters, respectively \citep{strong85}. Uncorrelated errors
can be computed by setting all off-diagonal elements to zero
before computing Eigenvectors and -values.

\end{document}